\documentclass[a4paper,11pt]{article}

\usepackage{jheppub} 
\usepackage{graphicx,color}
\usepackage{amsmath}
\usepackage{slashed}
\usepackage{multirow}
\usepackage{autobreak}
\usepackage[normalem]{ulem}
\usepackage{hyperref}
\usepackage{cleveref}
\usepackage{slashed}
\usepackage{amssymb}
\usepackage{epsfig}
\usepackage{array}
\usepackage{mathtools}
\usepackage{caption}
\usepackage[utf8]{inputenc}
\usepackage{mathtools, nccmath}
\usepackage{float}
\usepackage[section]{placeins}
\usepackage[utf8]{inputenc}
\usepackage[T1]{fontenc}
\usepackage{lmodern}
\usepackage[x11names]{xcolor}
\usepackage{framed}
\usepackage{tikz-feynman}
\colorlet{shadecolor}{blue!10}

\usepackage{array,booktabs}
\usepackage{subcaption}
\usepackage{contour}

\SetSymbolFont{largesymbols}{bold}{OMX}{txex}{b}{n}

\DeclareMathAlphabet{\mathpzc}{OT1}{pzc}{m}{it}

\newcommand{\be}{\begin{eqnarray*}}
	\newcommand{\ee}{\end{eqnarray*}}

\usepackage{parskip}

\newcommand{\ba}{\begin{array}}
	\newcommand{\ea}{\end{array}}
\newcommand{\bd}{\begin{displaymath}}
\newcommand{\ed}{\end{displaymath}}
\newcommand{\besub}{\begin{subequations}}
	\newcommand{\eesub}{\end{subequations}}

\def\q2 {q^2}

\def\bt{\begin{table}}
	\def\et{\end{table}}

\usepackage{bbm}  
\usepackage{amsmath}                                                

\newcommand{\nc}{\newcommand}
\nc{\beq}{\begin{equation}}  \nc{\eeq}{\end{equation}}
\nc{\bea}{\begin{eqnarray}}  \nc{\eea}{\end{eqnarray}}
\nc{\baa}{\begin{array}}     \nc{\eaa}{\end{array}}
\nc{\bit}{\begin{itemize}}   \nc{\eit}{\end{itemize}}
\nc{\ben}{\begin{enumerate}} \nc{\een}{\end{enumerate}}
\nc{\bce}{\begin{center}}    \nc{\ece}{\end{center}}
\nc{\bpm}{\begin{pmatrix}}   \nc{\epm}{\end{pmatrix}}
\nc{\bvt}{\begin{verbatim}}  \nc{\evt}{\end{verbatim}}
\nc{\bal}{\begin{align}}
\def\to{\rightarrow}

\def\boldoverdot{\,{\raise6pt\hbox{\bf.}\!\!\!\!\>}}

\def\ee{{\bf e}}

\usepackage{bm}
\def\diag{\hbox{\diag}}

\def\doubleundertext#1{
{\undertext{\vphantom{y}#1}}\par\nobreak\vskip-\the\baselineskip\vskip4pt%
\undertext{\hbox to 2in{}}}
\def\inbox#1{\vbox{\hrule\hbox{\vrule\kern5pt
     \vbox{\kern5pt#1\kern5pt}\kern5pt\vrule}\hrule}}
\def\sqr#1#2{{\vcenter{\hrule height.#2pt
      \hbox{\vrule width.#2pt height#1pt \kern#1pt
         \vrule width.#2pt}
      \hrule height.#2pt}}}
\def\square{\mathchoice\sqr56\sqr56\sqr{2.1}3\sqr{1.5}3}
\def\today{\ifcase\month\or
  January\or February\or March\or April\or May\or June\or
  July\or August\or September\or October\or November\or December\fi
  \space\number\day, \number\year}
\def\pmb#1{\setbox0=\hbox{#1}%
  \kern-.025em\copy0\kern-\wd0
  \kern.05em\copy0\kern-\wd0
  \kern-.025em\raise.0433em\box0 }

\def\pmbb#1{\setbox0=\hbox{#1}%
  \kern-.02em\copy0\kern-\wd0
  \kern.04em\copy0\kern-\wd0
  \kern-.02em\raise.03464em\box0 }
\def\sumprime_#1{\setbox0=\hbox{$\scriptstyle{#1}$}
  \setbox2=\hbox{$\displaystyle{\sum}$}
  \setbox4=\hbox{${}'\mathsurround=0pt$}
  \dimen0=.5\wd0 \advance\dimen0 by-.5\wd2
  \ifdim\dimen0>0pt
  \ifdim\dimen0>\wd4 \kern\wd4 \else\kern\dimen0\fi\fi
\mathop{{\sum}'}_{\kern-\wd4 #1}}
\title{\boldmath Probing flavor constrained SMEFT operators through $tc$ production at the Muon collider}

\author{Subhaditya Bhattacharya,}
\author{Sahabub Jahedi,}
\author{Soumitra Nandi, and}
\author{Abhik Sarkar}

\affiliation{Department of Physics, Indian Institute of Technology, Guwahati, Assam 781039, India}

\emailAdd{subhab@iitg.ac.in}
\emailAdd{sahabub@iitg.ac.in}
\emailAdd{soumitra.nandi@iitg.ac.in}
\emailAdd{sarkar.abhik@iitg.ac.in}

\abstract{We investigate flavor violating four-Fermi Standard Model Effective Field Theory (SMEFT) operators of dimension-six that can be probed via $tc ~(\bar{t}c+t\bar{c})$ production at the multi-TeV muon collider. We study different FCNC and FCCC processes related to $B$, $B_s$, $K$ and $D$ decays and mixings, sensitive to these operators and constrain the corresponding couplings. The tensor operator turns out to be most tightly bound. We perform event simulation of the final state signal from $tc$ production together with the SM background to show that operators after flavor constraint can reach the discovery limit at 10 TeV muon collider. We further adopt the optimal observable technique (OOT) to determine the optimal statistical sensitivity of the Wilson coefficients and compare them with the flavor constraints. We use the limits to predict the observational sensitivities of the rare processes like $K_L \to \pi_0 \ell \ell$, $D_0 \to \mu\mu$, $t \to c\ell\ell$.}


\DeclareUnicodeCharacter{2212}{-}

\begin{document} 
	
\maketitle
\flushbottom
	
\section{Introduction}
\label{sec:intro}
Search for physics beyond the Standard Model (BSM) at the Large Hadron Collider (LHC) and preceding experiments has not provided a conclusive result yet despite extensive search and data analysis. This suggests that BSM physics is either weakly coupled, or exists at energy scales considerably removed from 
	the electroweak scale, or possesses signals unaccounted from the standard searches done so far, or all of them. Therefore exploring BSM physics 
	via Standard Model Effective Field Theory (SMEFT) has gained considerable attention in recent times. The construction of effective theory assumes the 
	knowledge of only low scale theory (SM here), so that the Lagrangian involving 
	higher dimensional operators can be written as \cite{Weinberg:1979sa,Buchmuller:1985jz,Grzadkowski:2010es}:
	\beq
	\mathcal{L}=\mathcal{L}_{SM}+\sum^{\infty}_{n=5}\frac{1}{\Lambda^{n-4}}\sum_{i}C_i \mathcal{O}^{n}_{i},
	\eeq
	where $\mathcal{O}_i$'s are the effective operators constructed out of the Standard Model (SM) fields respecting the SM gauge symmetry, $C_i$'s are the new physics (NP) couplings or Wilson coefficients (WCs), through which the effect of BSM scenarios can be realised. $\Lambda$ denotes the NP scales integrated out, which has power ($n-4$) depending on the mass dimension $n$ of the operator $\mathcal{O}_i$ under consideration. SMEFT has been studied extensively via operators of dimension-five \cite{Weinberg:1979sa}, six \cite{Buchmuller:1985jz,Grzadkowski:2010es}, seven \cite{Lehman:2014jma,Bhattacharya:2015vja}, and eight \cite{Murphy:2020rsh,Li:2020gnx} to estimate NP deviations in a model-independent way. In this study, we will use the dimension-six SMEFT operators that contribute to $tc ~(\bar{t}c+t\bar{c})$ production at collider\footnote{Understandably the contributions from dimension-eight operators will be diminished by powers of $\Lambda^2$ and the effect will be milder, unless some specific assumption on NP is taken into account.} after taking the flavor constraints into account.  
	
	As we know that the renormalization-group equations (RGE) allow one to compute the running and mixing between the BSM scale down to the electroweak scale
	and further down to the scale of low-energy precision experiments, the SMEFT can be matched to a low energy EFT. It could be done following a two step 
	matching procedure. The coefficients $C_i$'s generated at the scale $\Lambda$ will be related to their values at the electroweak scale $v \sim 246$ GeV, and 
	these coefficients can further be related to the WCs of a low-energy EFT through RGE running. Therefore, any constraints obtained on the WCs from low 
	energy data will in turn limit $C_i(\Lambda)$'s at any given BSM scale $\Lambda$. The flavor changing neutral current (FCNC) processes are loop suppressed in 
	the SM, and any tree level NP contributions to such processes will have limited parameter spaces allowed by the data. 
	Similarly, the availability of precise data on the flavor changing charged current (FCCC) processes at the low energy could play an essential role to 
	constrain the NP parameter space contributing to such processes at the tree level. Therefore, study of the FCNC and FCCC processes 
	at the low energy will play a crucial role in exploring flavor violating NP effects under the SMEFT framework. 
	
	FCNC processes such as $b \to s(d)$ transitions are one of the most important probes in this connection, as there is no such interactions within the SM 
	at the tree level\footnote{Higher order corrections to $b \to s (d)$ transitions have been studied in \cite{Asatrian:2001de,Asatryan:2001zw,Asatrian:2003vq}.}. 
	Such flavor observables therefore provide an important bound on the corresponding SMEFT operators, see for 
	example, \cite{Aoude:2020dwv,Bissmann:2020mfi,Alda:2020okk,Bruggisser:2021duo,Grunwald:2023nli,Ali:2023kua}. 
	In this study, we focus on the observables related to the semi 
	leptonic and leptonic decays of $B$, $B_s$, $K$ mesons via the processes $b\to s (d) \ell^+\ell^-$ and $s\to d\ell^+\ell^-$ where the data is available only for 
	$\ell = \mu$ or $e$. In addition, we have considered the available data on $B-\bar{B}$ and $B_s-\bar{B_s}$ mixing amplitudes to constrain the NP parameter 
	spaces. Among the FCCC processes, we mainly consider the available data on semi leptonic and leptonic decays of $B$, $K$, $D$ and $D_s$ mesons via 
	$b\to c (u) \ell^+\bar{\nu}$, $s \to u \ell^+\bar{\nu}$ and $c \to d (s)\ell^+\bar{\nu}$ transitions, respectively.   
	
	Along with flavor observables, top-quark physics also plays an important role under SMEFT framework as it has $O(1)$ Yukawa coupling, which is 
	crucial to explain the origin of electro-weak symmetry breaking (EWSB). Several analyses under SMEFT framework in the top-quark sector have been 
	performed in \cite{Chala:2018agk,Bissmann:2019gfc,Altmannshofer:2023bfk}. This motivates us to choose $\bar{t}c (t\bar{c})$ production as an example 
	process to study. 
	
	The stringent constraint on the dimension-six effective operators derived from the flavor observables indicate that their production at the LHC is small, 
	while the SM background is huge and mostly irreducible, so that a prediction requires thorough analysis. If we choose the conservative limits,
	it is also difficult to probe them in a future electron-positron collider with a maximum centre-of-mass (CM) energy of 3 TeV. 
	Therefore we examine the process at high energy muon
	collider \cite{Black:2022cth}. Muons, being fundamental particles, provide the advantage of directing their entire energy towards short-distance scattering 
	rather than having it distributed among the partons. As a result, a 14 TeV muon collider can exhibit effectiveness on par with a 100 TeV proton-proton 
	collider \cite{Shiltsev:2019rfl}. This high-energy capability is particularly advantageous for both the exploration of new heavy particles along with 
	indirect measurements at elevated energy levels. Hence, a multi-TeV moun collider acts as both discovery and precision machine altogether. Limited 
	studies on $tc$ production at lepton colliders have been documented in the existing literature 
	\cite{Bar-Shalom:1997ezk,Bar-Shalom:1997htk,Bar-Shalom:1999dtk,Sun:2023cuf}. In a somewhat similar analysis \cite{Sun:2023cuf}, 
	the flavor constraints are obtained from rare B decay processes, applicable to the vector operators only, while ours is done considering all possible flavor 
	observables applicable to all of the vector, scalar and tensor type four-Fermi operators. 
	
	
	Further, we study the optimal statistical precision of the NP couplings using the optimal observable technique 
	(OOT) \cite{Atwood:1991ka,Davier:1992nw,Diehl:1993br,Gunion:1996vv}. OOT has been widely employed for precision analysis 
	in various domains, including Higgs couplings at  $e^+e^-$ colliders \cite{Hagiwara:2000tk,Dutta:2008bh}, top-quark properties at the
	$\gamma \gamma$ colliders \cite{Grzadkowski:2003tf,Grzadkowski:2004iw,Grzadkowski:2005ye} and $e^+e^-$ colliders \cite{Bhattacharya:2023mjr}, 
	CP properties of the Higgs at the LHC \cite{Gunion:1998hm}, $e \gamma$ colliders \cite{Cao:2006pu}, and muon collider \cite{Hioki:2007jc}. 
	Some recent works involving estimation of $Z$-couplings with heavy charged fermion \cite{Bhattacharya:2021ltd} and anomalous neutral triple gauge 
	couplings \cite{Jahedi:2022duc,Jahedi:2023myu} are also explored using the OOT.  
	
	The paper is arranged in the following way. In Section~\ref{sec:framework}, we discuss the relevant phenomenological framework for our analysis. 
	Then in Section~\ref{sec:flav}, we study the constraints from flavor physics, followed by collider simulation at $\mu^+\mu^-$ collider in Section~\ref{sec:col}. 
	Section \ref{sec:oot} elaborates OOT framework and the optimal limits on flavor violating dimension-six effective couplings. In Section~\ref{sec:scalar}, we discuss 
	the collider probe of scalar operator contributing to $3\to 2$ (hadronic) transition at $e^+e^-$ machine. Finally, we summarize our conclusion 
	in Section~\ref{sec:conclude}.
	
\section{Phenomenological framework}
\label{sec:framework}
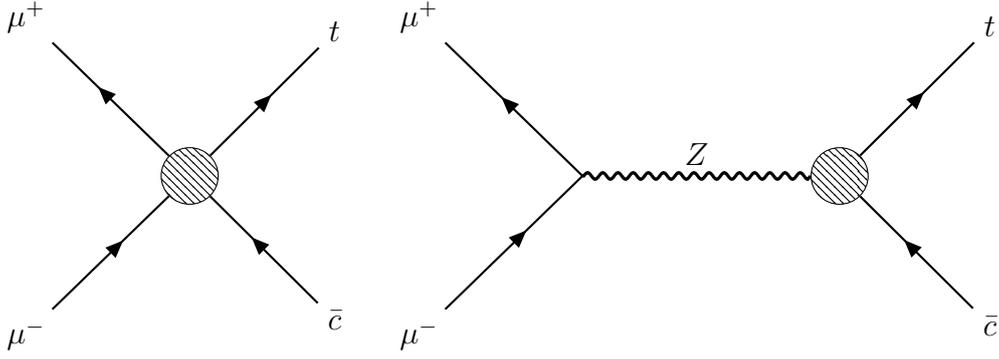
\begin{figure}[htb!]
		\centering
		\begin{tikzpicture}[baseline={(current bounding box.center),style={scale=0.7,transform shape}}]
		\begin{feynman}
		\vertex (a);
		\vertex [above left=2.5cm of a] (b) {\large $\mu^{+}$};
		\vertex [below left=2.5cm of a] (c) {\large $\mu^{-}$};
		\vertex[blob]  (d)  {\contour{black}{}};
		\vertex [above right=2.7cm of d] (e) {\large $t$};
		\vertex [below right=2.7cm of d] (f) {\large $\bar{c}$};
		\diagram{
			(b) -- [anti fermion, thick] (d) -- [ anti fermion, thick] (c);
			(d);
			(d) -- [fermion, thick] (e);
			(d) -- [ anti fermion, thick] (f);
		};
		\end{feynman}
		\end{tikzpicture} \quad
		\begin{tikzpicture}[baseline={(current bounding box.center),style={scale=0.7,transform shape}}]
		\begin{feynman}
		\vertex (a);
		\vertex [above left=2.5cm of a] (b) {\large $\mu^{+}$};
		\vertex [below left=2.5cm of a] (c) {\large $\mu^{-}$};
		\vertex[blob] [ right=3cm of a] (d)  {\contour{black}{}};
		\vertex [above right=2.8cm of d] (e) {\large $t$};
		\vertex [below right=2.8cm of d] (f) {\large $\bar{c}$};
		\diagram{
			(b) -- [anti fermion, thick] (a) -- [ anti fermion, thick] (c);
			(a) -- [ very thick, boson, edge label=\large  $Z$] (d);
			(d) -- [fermion, thick] (e);
			(d) -- [ anti fermion, thick] (f);
		};
		\end{feynman}
		\end{tikzpicture}	
		\caption{Feynman diagrams that induce $t \, \bar{c}$ production at the muon collider; left: effective four-Fermi ($\mu \mu tc$) couplings, 
			right: effective $Ztc$ couplings. The conjugate processes look alike with exactly same cross-section. }
		\label{fig:tcprod}
	\end{figure}

Our goal of the analysis is to probe the dimension-six flavor violating effective (EFT) operators that contributes to 
$tc ~(\bar{t}c+t\bar{c})$ production at the future muon collider. The Feynman graphs for the $t\bar{c}$ production are shown 
in figure~\ref{fig:tcprod} ($\bar{t}c$ graphs are similar with same contributions and not shown here). 
There are essentially two different contributions: $(i)$ via $Ztc$ vertex and $(ii)$ four-Fermi $\mu \mu tc$ contact interaction. 
As elaborated later, operators contributing to $Ztc$ affect the process very mildly, while the main contribution 
arises from the four-Fermi operators. 
	
	%
	
	There are three types of four-Fermi operators containing i) four leptons, ii) four quarks and iii) two leptons and two quarks, where the 
	last one serves our purpose. There are seven operators that contribute to the $\mu^+ \mu^- \to tc$ production, as given by 
	\cite{Bar-Shalom:1999dtk, Grzadkowski:2010es},
		\begin{align}
	\begin{split}
	\mathcal{O}^{(1)}_{l q}&=  (\bar{l}_p \gamma^{\mu} l_r)(\bar{q_i} \gamma_{\mu} q_j),\\
	\mathcal{O}^{(3)}_{l q}&=  (\bar{l}_p \gamma^{\mu} \tau^{I} l_r)(\bar{q_i} \gamma_{\mu}\tau^{I} q_j),\\
	\mathcal{O}_{eu}&= (\bar{e}_p \gamma^{\mu} e_r)(\bar{u}_i \gamma_{\mu} u_j),\\
	\mathcal{O}_{l u}&= (\bar{l}_p \gamma^{\mu} l_r)(\bar{u_i} \gamma_{\mu} u_j),\\
	\mathcal{O}_{qe}&=(\bar{q}_p \gamma^{\mu} q_r)(\bar{e}_i \gamma_{\mu} e_j),\\
	\mathcal{O}^{(1)}_{l e q u}&=  (\bar{l}^{a}_p e_r)\epsilon_{ab}(\bar{q}^{b}_i u_j),\\
	\mathcal{O}^{(3)}_{l e q u}&= (\bar{l}^{a}_p \sigma^{\mu \nu} e_r)\epsilon_{ab}(\bar{q}^{b}_i \sigma_{\mu \nu} u_j),\\
	\label{eq:ffops}
	\end{split}
	\end{align}
	where $l$ ($e$) is the left (right)-handed lepton doublet (singlet), $q$ is the left-handed quark doublet, and $u$ ($d$) is the up (down)-type right-handed quark 
	singlets. Apart, $p,r,i,j$ indicates flavor indices, $\tau^I$ are the Pauli matrices, $\epsilon=-i\tau_2$, $\sigma^{\mu\nu}=\frac{i}{2}[\gamma^\mu,\gamma^\nu]$. 
	Using Eq.~\eqref{eq:ffops}, the most general four-Fermi effective Lagrangian for $\mu\mu tc$ contact interaction can be written as,
	\beq
	\mathcal{L}_{\mu \mu tc}=\frac{1}{\Lambda^2}\sum_{s,t=L,R}\big[ \mathcal{V}_{st}(\bar{\mu}\gamma_{\mu}P_s \mu)(\bar{t}\gamma^{\mu}P_t c) + \mathcal{S}_{st}(\bar{\mu}P_s \mu)(\bar{t}P_t c) + \mathcal{T}_{st}(\bar{\mu}\sigma_{\mu \nu}P_s \mu)(\bar{t}\sigma^{\mu \nu}P_t c)\big],
	\label{eq:lageetc}
	\eeq 
	
	where vector-like ($\mathcal{V}_{st}$), scalar-like ($\mathcal{S}_{st}$) and tensor-like ($\mathcal{T}_{st}$) couplings can be expressed in terms of the WCs of 
	the seven four-Fermi operators as in Eq.~\eqref{eq:ffops} as,
	\begin{align}
	\begin{split}
	\mathcal{V}_{LL}&=(C^{(1)ij}_{l q}-C^{(3)ij}_{l q}), \quad \mathcal{V}_{LR}=C^{ij}_{l u}, \quad \mathcal{V}_{RR}=C^{ij}_{eu}, \quad \mathcal{V}_{RL}=C^{ij}_{qe}, \\ 
	\mathcal{S}_{RR}&=-C^{(1)ij}_{l equ}, \quad \mathcal{S}_{LL}=S_{LR}=S_{RL}=0,\\
	\mathcal{T}_{RR}&=-C^{(3)ij}_{l equ}, \quad \mathcal{T}_{LL}=\mathcal{T}_{LR}=\mathcal{T}_{RL}=0.\\
	\end{split}
	\end{align}
	Note that for vector operators we have all the helicity combination of fermions appear for both lepton and quarks, while for the scalar and tensor 
	couplings only $RR$ combination is non zero. In Eq.~\eqref{eq:lageetc}, the $\mathcal{S}_{LL}$ and $\mathcal{T}_{LL}$ also get a matching condition 
	from the hermitian conjugate SMEFT operators, leading to $-C^{(1)ji*}_{lequ}$ and $-C^{(3)ji*}_{lequ}$, respectively. The above parametrization will be used for the 
	collider simulation later. 
	
	The tree-level dimension-six EFT operators that contribute to effective {\it Ztc} vertex are, 
	\begin{align}
	\begin{split}
	\mathcal{O}^{(1)}_{\phi q}&=  i (\phi^{\dagger} D^{\mu} \phi)(\bar{q_i} \gamma_{\mu} q_j),\\
	\mathcal{O}^{(3)}_{\phi q}&=  i (\phi^{\dagger} D^{\mu} \tau^{I} \phi)(\bar{q_i} \gamma_{\mu} \tau^{I} q_j),\\
	\mathcal{O}_{\phi u}&= i (\phi^{\dagger} D^{\mu} \phi)(\bar{u_i} \gamma_{\mu} u_j).\\
	\end{split}
	\end{align}
	Therefore, the effective Lagrangian containing $Ztc$ vertex\footnote{$\gamma t c$ vertex is forbidden by $U(1)$ gauge invariance.} is,
	
	\beq
	\mathcal{L}_{Ztc}=g\frac{v^2}{\Lambda^2}\bar{t}\gamma^{\mu}(a^{Z}_L P_L+a^{Z}_R P_R)cZ_{\mu}\,,
	\eeq
	
	where, $P_{L(R)}=\frac{1-(+)\gamma^5}{2}$, $a_L$ and $a_R$ are the functions of $C_{\phi q}^{(1)ij}$, $C_{\phi q}^{(3)ij}$ and 
	$C^{ij}_{\phi u}$ as,
	
	\beq
	a^Z_L=\frac{1}{4 c_w}(C_{\phi q}^{(1)ij}-C_{\phi q}^{(3)ij})\, , \quad a^Z_L=\frac{1}{4 c_w} C^{ij}_{\phi u};
	\eeq
	
	where, $c_w=\cos \theta_w$, $\theta_w$ is weak mixing angle.
	
	
\section{Flavor constraints and predictions}
\label{sec:flav}
In this section, we study all possible flavor and top quark related experimental constraints that put bound on the dimension-six effective couplings of our interest as described above. We then make some observable predictions for some rare processes based on the constraints obtained for the effective operators.

\subsection{Flavor and top-quark constraints}
\label{sec:flav.cons}
Note that the operators defined in Eq.~\eqref{eq:ffops} will contribute to various low energy observables related to the FCNC and FCCC processes. In the following we will discuss different inputs which we have considered in this analysis.  

\paragraph{FCNC processes:} Most important constraints will be obtained from the semileptonic and leptonic decays of via $b\to s \ell^+\ell^-$ decays. The corresponding low energy effective Hamiltonian is given by \cite{Bobeth:1999mk, Altmannshofer:2008dz}  
\begin{align}
\nonumber
\mathcal H_\text{eff} =  \mathcal H_\text{eff}^\text{SM} - \frac{4G_F}{\sqrt{2}} \lambda^{bs}_t \frac{e_0^2}{16\pi^2} &\Big( \Delta C_9^\ell O_9^\ell + \Delta C_{10}^\ell O_{10}^\ell + C^{\prime \ell}_9 O^{\prime \ell}_9 + C^{\prime \ell}_{10} O^{\prime \ell}_{10} 
+ C_{S}^\ell O_S^\ell + C_P^\ell O_P^\ell \\
&+ C^{\prime \ell}_{S} O^{\prime \ell}_S + C_P^{\prime \ell} O^{\prime \ell}_P \Big),
\label{eq:bs.ham}
\end{align}
where $\ell$ stands here for all the three lepton fields: $e, \mu$ and $\tau$. Here, $e_0$ and $G_F$ are the $U(1)_{em}$ and Fermi coupling constants, $\lambda^{bs}_t=V_{tb}V_{ts}^{*}$ is the CKM combination. The four-Fermi operators as in Eq.~\eqref{eq:bs.ham} are expressed as,
\begin{align}
\begin{split}
O^{\ell}_9&=(\bar{s}\gamma_{\mu}P_L b)(\bar{\ell}\gamma^{\mu}\ell), \qquad O^{\ell}_S=(\bar{s}P_R b)(\bar{\ell}\ell)\,,\\
O^{\ell}_{10}&=(\bar{s}\gamma_{\mu}P_L b)(\bar{\ell}\gamma^{\mu} \gamma^{5}l), \quad O^{\ell}_P=(\bar{s}P_R b)(\bar{\ell}\gamma^{5}l)\,.\\
\end{split}
\label{eq:ops}
\end{align}	
The $O'$ operators can be obtained by interchanging $\{L \leftrightarrow R\}$ from the operators $O$ as above in Eq.~\eqref{eq:ops}. 
The SM Hamiltonian ($\mathcal{H}^{SM}$) already incorporates $O^{\ell}_9$ and $O^{\ell}_{10}$ operators with WCs $C^{\tt SM}_9$ and $C^{\tt SM}_{10}$. 
Therefore, $\Delta C^{\ell}_{9}$ and $\Delta C^{\ell}_{10}$ denote the WCs corresponding to $O^{\ell}_9$ and $O^{\ell}_{10}$ operators stemming from NP. 
Remaining effective operators are absent in the SM framework thus comes purely from NP contribution. We will use Eq.~\eqref{eq:bs.ham} and  Eq.~\eqref{eq:ops} to 
obtain limits from $b\to s$ and $b \to d$ transitions and translate them in terms of the WCs of our notation as in Eq.~\eqref{eq:ffops}. Following a tree level matching procedure, one can express the WCs defined in Eq.~\eqref{eq:ffops} in terms of the WCs of in the low energy effective Hamiltonian given above \cite{Aebischer:2015fzz}. 
\begin{itemize}
	\item \underline{$b \to s \mu \mu$ decay}:  $b \to s$ transitions are absent at the tree level within the SM; therefore observables 
	having such transitions provide intriguing indications of potential NP effects through $\Delta{C}^{\mu}_9$ and 
	$\Delta{C}^{\mu}_{10}$ \cite{Descotes-Genon:2012isb,Descotes-Genon:2013vna,Horgan:2013pva,Bharucha:2015bzk,Biswas:2020uaq}. 
	They stem from several experimental observations like $B \to K \mu \mu$, $B \to K^{*}\mu \mu$, and $B_s \to \phi \mu \mu$, which have been considered 
	in this analysis \cite{CDF:2011tds, LHCb:2013lvw,LHCb:2014cxe,LHCb:2014vgu,LHCb:2015svh,Belle:2016fev,CMS:2017rzx,ATLAS:2018gqc, LHCb:2020gog,LHCb:2021zwz}. The latest experimental results from the LHCb on lepton flavor university (LFU) ratio $R_K$ and $R^{*}_K$ \cite{LHCb:2022qnv,LHCb:2022vje} are 
	in great agreement with the SM prediction which indicates a tighter constraint on $\Delta{C}^{\mu}_9$ and $\Delta{C}^{\mu}_{10}$. 
	After matching dimension-six effective couplings in Eq.~\eqref{eq:ffops} with the Hamiltonian written in Eq.~\eqref{eq:bs.ham}, we get \cite{Aebischer:2015fzz},
	\begin{align}
	\begin{split}
	\Delta{C}_9^{\mu}=\left(\frac{\alpha}{\lambda^{bs}_t\pi}\right)\left(\frac{v^2}{\Lambda^2}\right)\left(C^{(1)32}_{lq}+C^{(3)32}_{lq}+{C}^{32}_{qe}\right), \\ 
	\Delta{C}_{10}^{\mu}=\left(\frac{\alpha}{\lambda^{bs}_t\pi}\right)\left(\frac{v^2}{\Lambda^2}\right)\left(C^{(1)32}_{lq}-C^{(3)32}_{lq}-{C}^{32}_{qe}\right), \\ 
	\end{split}
	\label{eq:bs.rel}
	\end{align}
	where $\alpha=e_0^2/4 \pi$ is fine-structure constant and $v$ is the vacuum expectation value (vev) of the SM Higgs. 
	We can see that $\mathcal{O}^{(1)}_{lq}$, $\mathcal{O}^{(3)}_{lq}$, and $\mathcal{O}_{qe}$ are the three dimension-six 
	SMEFT operators that contribute to $b \to s \mu \mu$ transition. Since we have analyzed the constraints by choosing one operator at a time, 
	therefore, for $\mathcal{O}_{qe}$  and $\mathcal{O}^{(3)}_{lq}$ operators, $\Delta{C}^{\ell}_9=-\Delta{C}^{\ell}_{10}$; 
	whereas for $\mathcal{O}^{(1)}_{lq}$, we have $\Delta{C}^{\ell}_9=\Delta{C}^{\ell}_{10}$. Following the analysis discussed in 
	\cite{Biswas:2020uaq}, we obtain $\Delta C_9^{\mu}=-0.23$ and transmitting this limit to couplings of dimension-six SMEFT operator using Eq.~\eqref{eq:bs.rel}, 
	the constraints are noted in Table~\ref{tab:cplng.const}. 
	
	\item \underline{$b \to d \mu \mu$ decay}: $b \to d \mu \mu$ transition is another FCNC process which also plays an important role to constrain 
	NP scenarios \cite{Rusov:2019ixr,Bause:2022rrs}. Like $b \to s \mu \mu$ transition, here also $\mathcal{O}^{(1)}_{lq}$, $\mathcal{O}^{(3)}_{lq}$, 
	and $\mathcal{O}_{qe}$ operators contribute. Consequently, relations between the WCs in two different parametrizations are similar to that 
	in Eq.~\eqref{eq:bs.rel}, just with a trivial substitution of $\lambda^{bd}_t=V_{tb}^{*}V_{td}$ and appropiate indices on WCs. We utilize the experimentally determined value of $\Delta{C}^{\mu}_9=-0.53$
	from the global fits of $B^+ \to \pi^+ \mu^+ \mu^-$, $B_s^0 \to \bar{K}^{*0} \mu^+ \mu^-$, $B^0 \to \mu^+ \mu^-$, and radiative $B \to X_d \gamma$ 
	decays data sourced from \cite{Bause:2022rrs}, to relate the dimension-six effective couplings of our choice (Eq.~\eqref{eq:bs.rel}). The constraints from this decay channel for these three dimension-six effective couplings are tabulated in Table~\ref{tab:cplng.const}.   
	
	\item \underline{$K_L \to \mu \mu$ decay}: The expression for the branching ratio of $K_L \to \mu^+ \mu^-$ decay using the 
	Hamiltonian in Eq.~\eqref{eq:bs.ham} can be expressed as \cite{Isidori:2003ts,Chobanova:2017rkj},
	\beq
	\text{BR}(K_L\to \mu^+\mu^-)=\frac{\tau_{L}f_K^2m_K^3\beta_{\mu}}{16\pi}\bigg|N^{LD}_{L}-\bigg(\frac{2m_{\mu}}{m_K}\bigg)\frac{G_F \alpha}{\sqrt{2}\pi}\text{Re}\bigg[-\lambda^{sd}_c\frac{Y_c}{s_W^2}+\lambda^{sd}_t C^{\mu}_{10}\bigg]\bigg|^2.
	\label{eq:klmm}
	\eeq
	In the above Eq.~\eqref{eq:klmm}, $\tau_L$ represents the mean life of the $K_L$ meson, $m_{\mu}$ and $m_K$ denote the masses of the muon and 
	kaon respectively, $G_F$ stands for the Fermi constant, and $\beta_{\mu}$ and $\lambda^{sd}_q$ 
	are defined as:
	\begin{align}
	\beta_{\mu}=\sqrt{1-\frac{4m_{\mu}^2}{m_K}}, \qquad \lambda^{sd}_q=V^{*}_{qd}V_{qs}.
	\end{align}
	$Y_c$ indicate short-distance contribution from SM while $N_{L}^{LD}=\pm[0.54(77)-3.95i]\times 10^{-11}$ $\rm{in~GeV^{-2}}$ is the long-distance 
	contribution \cite{Ecker:1991ru,DAmbrosio:2022kvb} and $C_{10}=C^{\tt SM}_{10}+\Delta{C}^{\mu}_{10}$. The relevant operators are $\mathcal{O}^{(1)}_{lq}$, 
	$\mathcal{O}^{(3)}_{lq}$ and $\mathcal{O}_{qe}$ that contribute to this observables. $\Delta{C}^{\mu}_{10}$ respects Eq.~\eqref{eq:bs.rel}. 
	Resulting limits on dimension-six effective couplings are written in Table~\ref{tab:cplng.const}.

	\item \underline{$B_q-\bar{B}_q$ mixing (with $q= d,s$):} $B^0$ or $B_s$ mesons exhibit oscillatory behaviour between particle and antiparticle states, 
	a phenomenon arising from the influence of flavor-changing weak interactions. These meson-antimeson mixings are the $\Delta B=2$ FCNC processes. 
	In presence of NP, the mixing amplitude or the frequency of meson-antimeson oscillations is defined as, 
	\begin{equation}
	\Delta M_q = 2 |M_{12}^q| = 2| M_{12}^{q,SM} +  M_{12}^{q,NP}|, 
	\end{equation}
	with
	\begin{equation}
	M_{12}^q = \frac{\langle B_q | \mathcal{H}_{eff}^{\Delta B=2}| \bar{B}_q \rangle }{2 M_{B_q}}\,.
	\end{equation}
	The dominant contribution to $M_{12}^{q,SM}$ in the SM arises from the top mediated box diagrams \cite{Hagelin:1981zk,Ali:1984db,Chau:1983te}, 
	and the corresponding contribution is given by, 
	\begin{equation}
	M_{12}^{q,SM} = \frac{G_F^2}{24 \pi^2}m_{B_q} f_{B_q}^2 \hat{B}_q \eta_{B} f(x^2) (\lambda_t^{bq})^2\,.
	\end{equation}  
	Here, $m_{B_d} (m_{B_s})$ and $f_{B^0_{d(s)}}=190.0~(230.3)$ MeV are the masses and decay constants, $\hat{B}_{d(s)}=1.222~(1.232)$ is the bag factor 
	of $B^0_{d(s)}$ meson respectively, $\eta_B=0.55$ is the QCD correlation factor \cite{FlavourLatticeAveragingGroupFLAG:2021npn}, and $\lambda^{bd(s)}_t=V_{tb}V_{td(s)}^*$. The Inami-Lim function $f(x)$ 
	is given by \cite{Gay:2000utx},
	\beq
	f(x)=x\left(\frac{1}{4}+\frac{9}{4(1-x)}-\frac{3}{2(1-x)^2}\right)-\frac{3x^3 \log x}{(1-x)^3},~~~~~x=\frac{m_t}{m_W}.
	\eeq
	
	\begin{figure}[t]
		$$
		\includegraphics[scale=0.25]{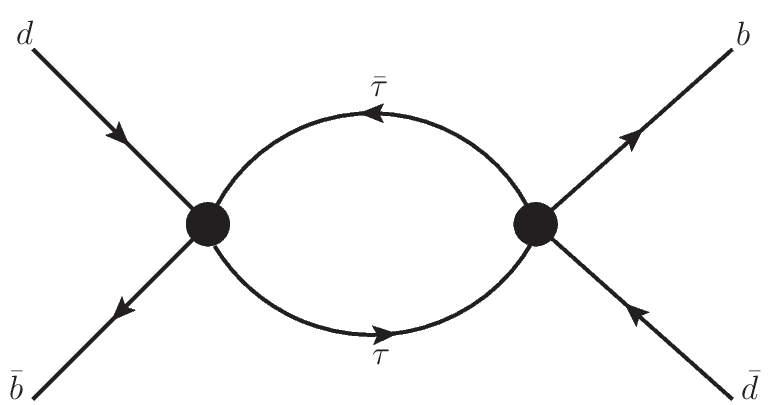}
		$$
		\caption{$\tau$ mediated 1-loop NP contribution to $\Delta m_{B^0_d}$.}
		\label{fig:1loop.mixing}
	\end{figure}
	
	We already mentioned that the SMEFT operators in Eq.~\eqref{eq:ffops} contribute to the FCNC processes like $b\to s\mu^+\mu^-$ and $b\to d \mu^+\mu^-$. 
	The four-Fermi operators in Eqs.~\eqref{eq:bs.ham} and \eqref{eq:ops}, may contribute to $B_s-\bar{B}_s$ and $B^0-\bar{B}^0$ mixing amplitudes respectively 
	via the diagram shown in figure \ref{fig:1loop.mixing}. In our scenario, the dispersive part of the diagram in figure~\ref{fig:1loop.mixing} will contribute to $M_{12}^q$. 
	The dominant contribution will arise from a $\tau$ mediated loop via the insertion of operators $\mathcal{O}_9$ and $\mathcal{O}_{10}$. From the 
	dispersive part of the diagram in figure~\ref{fig:1loop.mixing}, we obtain,
	\begin{align}
	M_{12}^{q,NP} & = -\frac{G_F^2 \alpha^2}{8 \pi^2} m_{B_i} f_{B_i}^2 \hat{B}_{i} \eta_{B} (\lambda^{bq}_t)^2\bigg[ 4\left((\Delta{C}_{9}^2+\Delta{C}_{10}^2) m_b^2 + 4\Delta{C}_{10}^2 m_{\tau}^2\right)\log\frac{\Lambda^2}{m_{\tau}^2} \nonumber \\ 
	& + (2m_b^2(\Delta{C}_9^2+\Delta{C}_{10}^2)+m_{\tau}^2(3\Delta{C}_{10}^2-\Delta{C}_9^2))
	\rm{DiscB[2m_b^2,m_{\tau},m_{\tau}]}\\\nonumber
	&+4 (\Delta{C}_9^2+\Delta{C}_{10}^2)m_b^2-2m_{\tau}^2 (\Delta{C}_9^2-7\Delta{C}_{10}^2)\bigg].
	\end{align}
	The expression of {\tt DisB[a,b,c]} function in the above equation is written in the appendix. As we have discussed earlier, the contributions to $b\to s \tau^+\tau^-$ will come from 
	$\mathcal{O}^{(1)}_{lq}$, $\mathcal{O}^{(3)}_{lq}$ and $\mathcal{O}_{qe}$.The amplitude is calculated using the Hamiltonian described in 
	Eq.~\eqref{eq:bs.ham}. Therefore, effective couplings related to these operators obey Eq.~\eqref{eq:bs.rel}. The relevant constraints from this 
	observable are presented in Table~\ref{tab:cplng.const}.

	\item \underline{$B^0_s \to \mu \mu$ decay}: Following Eq.~\eqref{eq:bs.ham}, the branching ratio of $B^0_s \to \mu^{+}\mu^{-}$ turns out \cite{Becirevic:2012fy},
	\beq
	\text{BR}(B^0_s \to \mu^{+} \mu^{-})=\tau_{B_s}\frac{G_F^2 \alpha^2 M_{B_s}f_{B_s}^2 m_{\mu}^2 (\lambda_{t}^{bs})^2}{16\pi^3}\sqrt{1-\frac{4 m_{\mu}^2}{M_{B_s}^2}}\big|C^{\mu}_{10}\big|^2.
	\eeq
	The branching ratio of this rare decay is $\left(2.69_{-0.35}^{+0.37}\right) \times 10^{-9}$ \cite{CMS:2014xfa,FlavourLatticeAveragingGroupFLAG:2021npn}.
	Understandably, $\mathcal{O}^{(1)}_{lq}$, $\mathcal{O}^{(3)}_{lq}$, and $\mathcal{O}_{qe}$ contribute to this branching ratio. Using Eq.~\eqref{eq:bs.rel}, 
	the constraints obtained on the effective couplings for the relevant operators are tabulated in Table~\ref{tab:cplng.const}.
	
	\item \underline{$K^\pm \to \pi^\pm \nu \nu$ decay:} The effective Hamiltonian relevant to $s \to d $ transitions is parameterized as \cite{Cirigliano:2011ny,Buchalla:1998ba},
	\beq
	\mathcal{H}^{s \to d}_{\tt eff}=\frac{G_F}{\sqrt{2}} \lambda^{sd}_t\left(\sum^{10}_{i=1}C_i O_i+\sum_{\ell =e, \mu, \tau} C^L_{\nu_{\ell}} O^L_{\nu_{\ell}}\right),
	\eeq
	where $\lambda^{sd}_t=V_{ts}V_{td}^*$. The details of  $O_1 - O_{10}$ operators are discussed in \cite{Cirigliano:2011ny}. The operator relevant to 
	$K^\pm \to \pi^{\pm} \nu \nu$ decay is
	\begin{align}
	\begin{split}
	O^L_{\nu_{\ell}}= \alpha (\bar{d} \gamma^{\mu} (1- \gamma^5) s)(\bar{\nu_{\ell}}\gamma_\mu(1-\gamma^5)\nu_{\ell}),\\
	O^R_{\nu_{\ell}}= \alpha (\bar{d} \gamma^{\mu} (1+ \gamma^5) s)(\bar{\nu_{\ell}}\gamma_\mu(1-\gamma^5)\nu_{\ell}).\\
	\end{split}
	\label{eq:sd.ops}
	\end{align} 
	The WC is $C^L_{\nu_{\ell}}=C^{L,\tt SM}_{\nu_{\ell}}+\delta C^L_{\nu_{\ell}}$ with SM contribution 
	\beq
	C^{L, \tt SM}_{\nu_{\ell}}=\frac{1}{2\pi s^2_w}\left(\frac{\lambda^{sd}_c}{\lambda^{sd}_t}X^{\ell}_c + X_t\right),
	\label{eq:sd.rel}
	\eeq
	where $X_t=1.48$ is the short-distance NLO QCD correction arises from top-quark, $X^e_{c} \simeq X^\mu_{c}=1.04 \times 10^{-3}$ and 
	$X_c^{\tau}=0.70 \times 10^{-3}$ are the short-distance charm contribution with the NLO logarithmic approximation \cite{Buchalla:1998ba,Buchalla:1995vs}. 
	By correlating the dimension-six effective operators from Eq.~\eqref{eq:ffops} with those specified in Eq.~\eqref{eq:sd.rel}, we can express the dimension-six effective 
	couplings \cite{Geng:2021fog} as,
	\beq
	\delta C^L_{\nu_{\ell}}=\frac{2 \pi}{e_0^2 \lambda^{sd}_t}\frac{v^2}{\Lambda^2}\left[C^{(1)21}_{l q}-C^{(3)21}_{l q}\right],~~~~ \qquad \delta C^R_{\nu_{\ell}}=0.
	\eeq
	The branching ratio of $B^+ \to \pi^+ \nu \nu$ decay is expressed as \cite{Geng:2021fog,Buras:2020xsm},
	\beq
	\mathcal{B}(K^+\to\pi^+\nu\bar{\nu})=\frac{2\alpha^2|\lambda^{sd}_t|^2 \mathcal{B}(K^+\to\pi^0 e^+\nu_e)}{|V_{us}|^2}\sum_{\ell=e,\mu,\tau}|C_{\nu_\ell}^L+C_{\nu_\ell}^R|^2\,.
	\label{eq:k.pinunu}
	\eeq
	We take $\mathcal{B}(K^+\to\pi^0 e^+\nu_e)=(5.07\pm0.04)\times10^{-2}$ and $\mathcal{B}(K^+\to\pi^+\nu \nu)=(1.14_{-0.33}^{+0.40})\times10^{-10}$ from the PDG average~\cite{ParticleDataGroup:2022pth}, and obtain $\delta C_{\nu_{\ell}}=-3.31$. Following Eq.~\eqref{eq:sd.rel}, constraints on dimension-six effective couplings are 
	noted in Table~\ref{tab:cplng.const}.
	
\end{itemize}

	\begin{table}[htb!]
	{\renewcommand{\arraystretch}{1.3}%
		\centering
		\scalebox{0.9}{
			\begin{tabular}{ |c|c|c|c|c| } 
				\hline
				NP  & Related processes/ & Constraints  & Simultaneous fit \\ 
				couplings& Observables &($\rm{GeV^{-2}}$)& ($\rm{GeV^{-2}}$)\\
				\hline
				&  $b\to s \mu \mu $   & $-(3.98\pm 1.20)\times10^{-10}$ &  \\ 
				$C^{(1)32}_{lq}/\Lambda^2$&$\Delta{m_{B_s^0}}$&$-(3.08\pm0.01)\times10^{-8}$&$-(4.94\pm0.03)\times10^{-9}$\\
				& $\mathcal{B}(B^0_s \to \mu^+\mu^-)$ & $-(7.39_{-0.44}^{+0.41})\times10^{-10}$ &\\
				\hline
				\multirow{2}{*}{$C^{(1)31}_{lq}/\Lambda^2$}& $b\to d \mu \mu $ & $-(1.90_{-0.68}^{+0.65})\times10^{-10}$ \cite{Bause:2022rrs} &\multirow{2}{*}{$-(5.01 \pm 0.66) \times 10^{-10}$}\\ 
				&$\Delta{m_{B_d^0}}$&$-(1.81\pm0.05)\times10^{-8}$&\\
				\hline
				\multirow{2}{*}{$C^{(1)21}_{lq}/\Lambda^2$} & $\mathcal{B}(K_{L} \to \mu^+\mu^-)$ & $-(1.21_{-0.31}^{+0.33})\times10^{-10}$ &\multirow{2}{*}{$-(1.21\pm0.32)\times10^{-10}$}\\
				& $\mathcal{B}(K^+ \to \pi^+  \nu \bar{\nu})$ & $-(3.50_{-0.80}^{+0.55})\times10^{-10}$ &\\ 
				\hline
				\hline
				&$B \to D^{(*)}\tau \nu_{\tau}$& $-(0.35\pm0.15) \times 10^{-7}$ \cite{Ray:2023xjn}&\\
				&$B \to D^{(*)}\mu(e) \nu$& $(0.14\pm0.09) \times 10^{-7}$ \cite{Ray:2023xjn}&\\
				$C^{(3)32}_{lq}/\Lambda^2$& $b\to s \mu \mu $   & $-(3.98\pm 1.20)\times10^{-10}$ & $-(2.15\pm 0.04)\times10^{-9}$  \\ 
				& $\mathcal{B}(B^0_s \to \mu^+\mu^-)$ & $-(7.39_{-0.44}^{+0.41})\times10^{-10}$ &\\
				&$\Delta{m_{B_s^0}}$&$-(1.10\pm0.01)\times10^{-8}$&\\
				\hline
				&$B \to \pi\mu \nu_{\mu}$&  $(0.06\pm0.20) \times 10^{-7}$ \cite{Ray:2023xjn}&\\
				\multirow{2}{*}{$C^{(3)31}_{lq}/\Lambda^2$}& $b\to d \mu \mu $ & $-(1.90_{-0.68}^{+0.65})\times10^{-10}$ \cite{Bause:2022rrs} &\multirow{2}{*}{$-(5.01 \pm 0.66) \times 10^{-10}$}\\ 
				&$\mathcal{B}(B^{\pm}\to \tau \nu_{\tau})$& $(1.27\pm0.43) \times 10^{-7}$&\\
				&$\Delta{m_{B_d^0}}$&$-(1.81\pm0.05)\times10^{-8}$&\\
				\hline
				& $\mathcal{B}(K_{L} \to \mu^+\mu^-)$ & $-(1.21_{-0.31}^{+0.33})\times10^{-10}$ &\\
				\multirow{2}{*}{$C^{(3)21}_{lq}/\Lambda^2$}&$\mathcal{B}(K^{+}\to \pi^+ \nu \bar{\nu})$&$(3.50_{-0.80}^{+0.55})\times10^{-10}$&\multirow{2}{*}{$(1.63 \pm 0.29) \times 10^{-10}$}\\ 
				&$\mathcal{B}(K^{\pm}\to \mu \nu_{\mu})$& $(2.21\pm0.74) \times 10^{-8}$&\\
				&$\mathcal{B}(D^{\pm}\to \mu \nu_{\mu})$& $-(1.14\pm1.52) \times 10^{-8}$&\\
				\hline
				$C^{(3)11}_{lq}/\Lambda^2$&$\mathcal{B}(\pi^{\pm} \to \mu \nu_{\mu})$& $(1.28\pm0.81) \times 10^{-8}$& \\
				\hline
				$C^{(3)22}_{lq}/\Lambda^2$&$\mathcal{B}(D_s^{\pm}\to \mu \nu_{\mu})$& $(6.02\pm9.94) \times 10^{-9}$&\\
				\hline
				$C^{(3)33}_{lq}/\Lambda^2$& $\mathcal{B}(t\to b \ell \nu_{\ell})$ & $(7.13\pm0.37)\times10^{-7}$&\\
				\hline\hline
				$C^{32}_{eu}/\Lambda^2$& $t \to c \ell \ell$ & $<10^{-5}$ \cite{Altmannshofer:2023bfk}&-\\
				\hline\hline
				$C^{32}_{lu}/\Lambda^2$& $ t  \to c \ell \ell$ & $<10^{-5}$ \cite{Altmannshofer:2023bfk}&-\\
				\hline\hline
				& $b\to s \mu \mu $   & $-(1.24\pm 0.26)\times10^{-9}$ &  \\ 
				$C^{32}_{qe}/\Lambda^2$& $\mathcal{B}(B^0_s \to \mu^+\mu^-)$ & $-(7.39_{-0.445}^{+0.415})\times10^{-10}$ &$-(1.01 \pm 0.02)\times 10^{-9}$\\
				&$\Delta{m_{B_s^0}}$&$-(1.10\pm0.01)\times10^{-8}$&\\
				\hline
				\multirow{2}{*}{$C^{31}_{qe}/\Lambda^2$}& $b\to d \mu \mu $ & $(0.68_{-2.79}^{+2.72})\times10^{-10}$ \cite{Bause:2022rrs} &\multirow{2}{*}{$-(4.15 \pm 0.24)\times10^{-9}$}\\ 
				&$\Delta{m_{B_d^0}}$&$-(1.81\pm0.05)\times10^{-8}$&\\
				\hline
				$C^{21}_{qe}/\Lambda^2$ & $\mathcal{B}(K_{L} \to \mu^+\mu^-)$ & $-(1.213_{-0.310}^{+0.334})\times10^{-10}$ &\\
				\hline
		\end{tabular}}
		\caption{Constraints on the different dimension-six effective vector couplings from various flavor and top-quark processes and decay modes. For the details 
			on observables concerning $b\to s \mu^+\mu^-$ and $b\to d\mu^+\mu^-$ transitions, see references \cite{Biswas:2020uaq,Bause:2022rrs} and 
			for the measurements see the text.}
		\label{tab:cplng.const}}
\end{table} 

\begin{table}[t]
	{\renewcommand{\arraystretch}{1.3}%
		\centering
		\scalebox{0.9}{
			\begin{tabular}{ |c|c|c|c|c| } 
				\hline
				NP  & Related processes/ & Constraints  & Simultaneous fit \\ 
				couplings& Observables &($\rm{GeV^{-2}}$)& ($\rm{GeV^{-2}}$)\\
				\hline
				&$B \to D^{(*)}\tau \nu_{\tau}$&  $(1.81\pm0.06) \times 10^{-7}$ \cite{Ray:2023xjn}&\\
				\cline{2-3} 
				$C^{(1)32}_{l e q u}/\Lambda^2$&$B\to D^{(*)} \mu \nu_{\mu}$ & \multirow{2}{*}{$(0.09\pm0.58) \times 10^{-7}$ \cite{Ray:2023xjn} } &$(1.79\pm0.06)\times 10^{-7}$\\
				&$B\to D^{(*)} e \nu_{e}$ &  & \\
				\hline
				&$B\to \pi \mu \nu_{\mu}$& \multirow{2}{*}{$(0.03\pm0.18) \times 10^{-7}$ \cite{Ray:2023xjn}} &\\
				$C^{(1)31}_{l e q u}/\Lambda^2$&$B\to \pi e \nu_{e}$&  &$-(6.64\pm0.21)\times 10^{-9}$\\
				\cline{2-3}
				&$\mathcal{B}(B^{\pm}\to \tau \nu_{\tau})$& $-(6.64\pm0.21) \times 10^{-9}$&\\
				\hline
				\multirow{2}{*}{$C^{(1)21}_{l e q u}/\Lambda^2$}&$\mathcal{B}(D^{\pm}\to \mu \nu_{\mu})$& $-(8.61\pm0.18) \times 10^{-10}$&\multirow{2}{*}{$-(9.72\pm 0.17) \times 10^{-10}$}\\
				&$\mathcal{B}(K^{\pm}\to \mu \nu_{\mu})$& $-(1.83\pm0.05) \times 10^{-9}$&\\
				\hline
				 $C^{(1)11}_{l e q u}/\Lambda^2$&$\mathcal{B}(\pi^{\pm} \to \mu \nu_{\mu})$&  $-(5.82\pm0.31) \times 10^{-10}$&  \\
				 \hline
				$C^{(1)22}_{l e q u}/\Lambda^2$&$\mathcal{B}(D_s^{\pm}\to \mu \nu_{\mu})$& $-(4.36\pm3.17) \times 10^{-10}$&\\
				\hline\hline
				&$B \to D^{(*)}\tau \nu_{\tau}$&  $(0.62\pm0.17) \times 10^{-7}$ \cite{Ray:2023xjn}&\\
				\cline{2-3}
				$C^{(3)32}_{l e q u}/\Lambda^2$& $B\to D^{(*)} \mu \nu_{\mu}$ & \multirow{2}{*}{$-(0.33\pm 6.61) \times 10^{-10}$ \cite{Ray:2023xjn}} & $(0.61\pm 6.60)\times10^{-10}$ \\
				 & $B\to D^{(*)} e \nu_{e}$ &  &  \\
				\hline
				\multirow{2}{*}{$C^{(3)31}_{l e q u}/\Lambda^2$}&$B\to \pi \mu \nu_{\mu}$&  \multirow{2}{*}{$(0.07\pm0.16) \times 10^{-7}$ \cite{Ray:2023xjn}}&\\
				& $B\to \pi e \nu_{e}$ &  & \\
				\hline
		\end{tabular}}
		\caption{Constraints of scalar and tensor mediated dimension-six effective couplings from various flavor observables, see text for details.}
		\label{tab:cplng.const2}}
\end{table} 

\paragraph{FCCC processes:}

\begin{itemize}
	\item \underline{$b \to q \ell \nu_{\ell}$ decay}: The most general effective Hamiltonian for $ b \to  c(u) \ell \nu_{\ell} $ transitions is written as \cite{Sakaki:2013bfa},
	\beq
	\mathcal{H}^{b\to q \ell \bar{\nu}_{\ell}}_{\text{eff}}=\frac{4G_F}{\sqrt{2}}V_{qb}[(1 + C_{V_1}^{\ell})\mathcal{O}_{V_1}^{\ell}+C_{V_2}^{\ell} \mathcal{O}_{V_2}^{\ell}+C_{S_1}^{\ell}\mathcal{O}_{S_1}^{\ell}+C_{S_2}^{\ell}\mathcal{O}_{S_2}^{\ell}+C_T^{\ell} \mathcal{O}_T^{\ell}] \,,
	\label{eq:bc.ham}
	\eeq
	with the operators
	\begin{align}
	\begin{split}
	\mathcal{O}_{V_1}^{\ell} &= (\bar{q}_L\gamma^\mu b_L)(\bar{\ell}_L\gamma_\mu\nu_{\ell L}),\\
	\mathcal{O}_{V_2}^{\ell} &= (\bar{q}_R\gamma^\mu b_R)(\bar{\ell}_L\gamma_\mu\nu_{\ell L}),\\
	\mathcal{O}_{S_1}^{\ell} &=(\bar{q}_L b_R)(\bar{l}_R\nu_{\ell L}),\\
	\mathcal{O}_{S_2}^{\ell} &= (\bar{q}_R b_L)(\bar{l}_R\nu_{\ell L}),\\
	\mathcal{O}_{T}^{\ell} &= (\bar{q}_R\sigma^{\mu\nu}b_L)(\bar{\ell}_R\sigma_{\mu\nu}\nu_{lL}),\\
	\end{split}
	\label{eq:bc.ops}
	\end{align}
	where $q = u$ or $c$. If we match the ones in Eq.~\eqref{eq:bc.ham} with those mentioned in Eq.~\eqref{eq:ffops} for $b \to c \ell \nu_{\ell}$ transition, we find the following relations\footnote{For $b \to u \ell \nu_{\ell}$ transition, $c$ index in the CKM matrix elements is replaced by $u$ and the WCs will have `31' flavor indices.} 
	among the couplings \cite{Aebischer:2015fzz},
	\beq
	\frac{C^{(3)32}_{lq}}{\Lambda^2}=-\frac{V_{cb}}{V_{cs}}\left(\frac{C^{\ell}_{V_1}}{v^2}\right),~~~\frac{C^{(1)32}_{lequ}}{\Lambda^2}=-2\frac{V_{cb}}{V_{tb}}\left(\frac{C^{\ell}_{S_2}}{v^2}\right),~~~\frac{C^{(3)32}_{lequ}}{\Lambda^2}=-2\frac{V_{cb}}{V_{tb}}\left(\frac{C^{\ell}_{T}}{v^2}\right).
	\label{eq:bc.rel}
	\eeq
	FCCC processes like $b \to q \ell \nu$ have been widely studied on \cite{Bhattacharya:2018kig,Huang:2018nnq,Ray:2023xjn,Fedele:2023ewe}. 
	In this analysis, we borrow the constraints from \cite{Ray:2023xjn}, where the experimental results of $\bar{B} \to D(D^*)\ell^-\bar{\nu}_{\ell}$, 
	$R(D^{(*)})$,$\bar{B} \to \pi \ell^- \bar{\nu}_{\ell}$ ($\ell=\mu,~e$) 
	along with lattice inputs are taken into account. We find that from $b \to c \mu \nu_{\mu}$ and $b \to u \mu \nu_{\mu}$ decays, the constraints on 
	$\mathcal{O}^{(1)}_{l e q u}$ are $C^{(1)32}_{l e q u}/\Lambda^2=(0.09 \pm 0.58)\times10^{-7}$ $\rm{GeV^{-2}}$ and 
	$C^{(1)31}_{lequ}/\Lambda^2=(0.02 \pm 0.12)\times10^{-7}$ $\rm{GeV^{-2}}$ respectively. As the uncertainties are very high here, we may infer that
	the corresponding operator is insensitive to these decays and zero consistent. $\mathcal{O}^{(3)}_{l e q u}$ is also insensitive for these two decay modes 
	for the same reason.
	
	\item \underline{$P \to \ell \nu_{\ell}$ decay}: Leptonic decays from pseudoscalar mesons ($P$) provide one of cleanest probes to constrain NP. 
	The hadronic matrix elements for these decays with different Lorentz structures are defined as \cite{Becirevic:2020rzi},
	\beq
	\left \langle 0|\bar{q}_1 \gamma^{\mu} \gamma^5 q_2|P(p) \right \rangle =i f_P p^{\mu}, ~~~~~~\left \langle 0|\bar{q}_1  \gamma^5 q_2|P(p) \right \rangle=-i f_P \frac{M_P}{m_{q_1}+m_{q_2}}\,,
	\eeq 
	where $m_P$, $f_P$, and  $m_{q_{1(2)}}$ are the mass, decay constant, and mass of constituent quarks of $P$-meson, resepectively. These inputs for different 
	pseudoscalar mesons  are taken from the FLAG review \cite{FlavourLatticeAveragingGroupFLAG:2021npn}.
	$\mathcal{O}^{(3)}_{l q}$ and $\mathcal{O}^{(1)}_{l equ}$ operators contribute to these decays. Using the effective Hamiltonian in Eq.~\eqref{eq:bc.ham}, 
	the branching ratio of $P \to \ell \nu_{\ell}$ is expressed as;
	\beq
	\mathcal{B}(P \to \ell \nu_{\ell})=\tau_P \frac{m_P m_{\ell}f^2_P G^2_F V_{q_1 q_2}}{8 \pi}\left(1-\frac{m_{\ell}^2}{m_P^2}\right)^2\left|1-C_{V_1}+\frac{m_P^2}{m_{\ell}(m_{q_1}+m_{q_2})}C_{S_2}\right|^2\,,
	\eeq
	where, $\tau_P$ is the lifetime of $P$ meson, $m_{\ell}$ is the mass of the lepton ($\ell$), and $V_{q_1 q_2}$ is CKM matrix element. 
	$C_{V_1}$ and $C_{S_2}$ follow same relations  as expressed in Eq.~\eqref{eq:bc.rel}. Constraints on $C_{V_1}$ and $C_{S_2}$ couplings are tabulated 
	in Table~\ref{tab:sclr.coup}. Using these observables, we note the constraints on $C_i/\Lambda^2$ in Table~\ref{tab:cplng.const} and \ref{tab:cplng.const2} 
	following Eq.~\eqref{eq:bc.rel}.
	
	\begin{table}[t]
		\centering
		{\renewcommand{\arraystretch}{1.4}%
			\begin{tabular}{| c |c |c| c| }
				\hline
				\multicolumn{1}{|c}{Decay }&
				\multicolumn{1}{|c}{ Branching}&
				\multicolumn{2}{|c|}{Constraints $\rm (GeV^{-2})$}\\
				\cline{3-4}
				modes & ratio & $C_{V_1}$ &$C_{S_2}$ \\ 
				\hline
				$\pi^{\pm}\to \mu \nu_{\mu}$ & $(99.9877 \pm 0.00004) \times 10^{-2}$ & $-(1.77 \pm 0.6) \times 10^{-2}$  & $(4.33\pm0.23)\times10^{-4}$  \\  
				$K^{\pm}\to \mu \nu_{\mu}$ & $(63.56 \pm 0.11)\times 10^{-2}$ &  $-(3.21 \pm 1.08) \times 10^{-2}$ & $(1.36\pm 0.04)\times10^{-3}$  \\
				$B^{\pm}\to \tau \nu_{\tau}$ & $(1.09 \pm 0.24) \times 10^{-4}$ & $-(1.85 \pm 0.62)\times 10^{-1}$ &$(4.94\pm1.60)\times10^{-2}$ \\
				$D_s^{\pm}\to \mu \nu_{\mu}$ & $(5.43 \pm 0.15) \times 10^{-3}$ &$-(8.74\pm14.43)\times10^{-3}$ &$(3.25\pm2.36)\times10^{-4}$ \\
				$D^{\pm}\to \mu \nu_{\mu}$ & $(3.74 \pm 0.17) \times 10^{-4}$ & $(1.66 \pm 2.21) \times 10^{-2}$ &$-(6.40\pm8.71)\times10^{-4}$ \\
				\hline
		\end{tabular}}
		\caption{Constraints on $C_{V_1}$ and $C_{S_2}$ from $\mathcal{B}$($P \to \ell \nu_{\ell}$). Experimental inputs for different $P \to \ell \nu_{\ell}$ branching ratios are adopted from the PDG average \cite{ParticleDataGroup:2022pth}.}
		\label{tab:sclr.coup}
	\end{table}
	
	\item  \underline{$t \to b \ell \nu_{\ell}$ decay}: Along with the flavor observable described above, semi leptonic decays of top quark is instrumental in constraining NP. The $t \to b \ell \nu_{\ell}$ decay gets contribution from few operators in Eq.~\eqref{eq:ffops}, thereby imposing constraints on the associated dimension six effective couplings. This decay process is governed by $W$ mediated charge current interaction within SM whereas for NP contribution 
	arises from $\mathcal{O}^{(3)}_{lq}$, $\mathcal{O}^{(1)}_{lequ}$, and $\mathcal{O}^{(3)}_{lequ}$ operators. The total decay width ($\Gamma^{t}$) for this process is given by,
	\beq
	\Gamma^t=\frac{1}{64 \pi^3 m_t} \int_{E_{\ell}=0}^{\frac{m_t}{2}} \left( \int^{\frac{m_t}{2}}_{E_b=(\frac{m_t}{2} - E_{\ell})} \left|\mathcal{M}^{t}_{\tt tot}\right|^2 dE_b \right) dE_{\ell},
	\eeq
	where $m_t$ is mass of top quark, $E_b$ and $E_{\ell}$ are the energy of b-quark and lepton ($\ell$), respectively. SM and NP contribution to the total amplitude are written in the appendix. Constraints on  $\mathcal{O}^{(3)}_{lq}$ from this decay is tabulated in Table~\ref{tab:cplng.const}.  For scalar ($C^{(3)}_{\ell e q u}/\Lambda^2$) and tensor ($C^{(3)}_{\ell e q u}/\Lambda^2$) mediated couplings, orders of the constraints are $10^{-3}$ and $10^{-4}$ respectively. For $t \to c \ell \ell$ decay (Eq.~\eqref{eq:tcll}), upper limit of these operstors is $10^{-5}$. Therefore constraints obtained from $ t \to b \ell \nu$ on $C^{(1)}_{\ell e q u}/\Lambda^2$ and $C^{(3)}_{\ell e q u}/\Lambda^2$ are ruled out from $t \to c \ell \ell$ decay.

	
\end{itemize}
 As our interest lies mostly in $3 \to 2$ transitions, WCs of the contributing operators are more constrained from 
present data involving electron or muon final states compared to tau. However, both electron and muon 
final states provide similar bounds. Therefore, we do not consider lepton generation indices in the WCs and add indices pertaining 
to quark sectors only. Utilizing experimental data from various flavor observables and top quark measurements, individual 
constraints on different WCs are presented in column three of Table~\ref{tab:cplng.const} and \ref{tab:cplng.const2}. Simultaneous fit is performed to 
constrain the WCs, incorporating all available flavor violating observables pertaining to specific $i\to j$ transitions, see column four of 
Table~\ref{tab:cplng.const} and \ref{tab:cplng.const2}. If we further perform 
a simultaneous fit by considering all the flavor violating contributions, then the constraints on vector, scalar, and tensor couplings turn out to be:
\begin{align} 
\begin{split}
\nonumber
C^{(1)32}_{l q}/\Lambda^2&=-9.96 \times 10^{-10} ~{\rm{GeV^{-2}}}, \\ C^{(1)32}_{lequ}/\Lambda^2&=-9.27 \times 10^{-10}~{\rm{GeV^{-2}}},  \\
C^{(3)32}_{lequ}/\Lambda^2&=7.24 \times 10^{-11}~\rm{GeV^{-2}}. 
\end{split}
\end{align}
These bounds are of similar order for vector and scalar couplings but more stringent for the tensor one, all of which we probe at the muon collider (Section \ref{sec:col}). 
We would like to note however, that the scalar coupling from $3 \to 2$ transition 
are less constrained, $C_{lequ}^{(1)32}/\Lambda^{2} \sim 10^{-7}$ GeV$^{-2}$, see in Table \ref{tab:cplng.const2}. 
Therefore, it can be individually proved with less CM energy (preferably at $e^+e^-$ colliders) which we discuss in Section \ref{sec:scalar}. 
However, with more data for $3 \to 2$ transition in future, $C^{(1)32}_{lequ}/\Lambda^2$ constraints could reach above mentioned simultaneous fit bound.  
Note that no suitable observable has been found to constrain $\mathcal{O}_{l u}$ and $\mathcal{O}_{e u}$ operators, resulting an upper bound on them from 
$ t \to c \ell \ell$ decay, as pointed out in Table~\ref{tab:cplng.const}. We make predictions for $ t \to c \ell \ell$ transition and other flavor observables as we discuss next.
%
\subsection{Prediction on different observables}
\label{sec:pred}
Using the constraints on the EFT operators obtained in Table~\ref{tab:cplng.const}, Table~\ref{tab:cplng.const2} and Table \ref{tab:sclr.coup}, 
we provide future prediction of following observables, with numerical estimates noted in Table~\ref{tab:br.prdctn}.
\begin{itemize}
	\item \underline{$\mathcal{B}$($t\to c \ell \ell$)}: The branching ratio of $t \to c \ell \ell$ decay is given 
	by\footnote{The expression of Br($t\to c \ell \ell$) is calculated in the limit of $m_c,m_{\ell} \to 0$, 
		but in numerical evaluation all masses are taken into account.},
	\begin{align}
		\begin{split}
	\mathcal{B}(t \to c \ell \ell)=\frac{m_t^5}{6144 \pi^3 \Lambda^4 \Gamma_{\tt tot}^{t}}&\bigg(4(C^{(1)32}_{l q}+C^{(3)32}_{l q})^2+\big(C^{32^2}_{eu}+C^{32^2}_{l u}+C^{32^2}_{qe}\\
	&+C^{(1)32^2}_{l e q u}+72C^{(3)32^2}_{l e q u}\big)\bigg)\,.
	\label{eq:tcll}
	\end{split}
	\end{align}
	As of now, there hasn't been any specific experimental searches conducted to observe $t \to c \ell \ell$ decays. However, an indirect upper limit on 
	$t \to c \ell \ell$ decays can be obtained by $t \to Zq$ searches at ATLAS \cite{ATLAS:2018zsq}. At 95\% C.L., bounds on $t \to c \ell \ell$ branching 
	ratios are \cite{Altmannshofer:2023bfk},
	\begin{align}
	\begin{split}
	\label{eq:tcll.bnd}
	\mathcal{B}(t \to c e^+ e^-) &< 2.1 \times 10^{-4},\\
	\mathcal{B}(t \to c \mu^+ \mu^-)&< 1.5 \times 10^{-4}.\\
	\end{split}
	\end{align}
	These upper limits of the branching ratio provide upper bounds on  the dimension-six effective vector and scalar operators on the order of $10^{-5}$ whereas for tensor operator the upper bound is $10^{-7}$. Constraints determined from experimental inputs on dimension-six effective couplings listed in Table~\ref{tab:cplng.const} and \ref{tab:cplng.const2} respect this upper bound. Using the obtained constraints on dimension-six effective couplings, we provide the prediction on this observable in Table \ref{tab:br.prdctn}.
	
	\item \underline{$\mathcal{B}$($K_{L}\to \pi^0 \ell \bar{\ell}$)}: The branching fraction of $K_{L}\to \pi^0 \ell \bar{\ell}$ can be expressed as \cite{Ecker:1987hd,Ecker:1987qi},
	\begin{align}
	\mathcal{B}(K_L \to \pi^0 \ell \bar{\ell}) = \left( C_{\rm dir}^\ell \pm C_{\rm int}^\ell|a_S| + C_{\rm mix}^\ell|a_S|^2 + C_{\gamma \gamma}^\ell  \right)\cdot 10^{-12}\,,
	\label{eq:klp0}
	\end{align}
	The details of $a_s$, $C^{\ell}_{\tt dir}$, $C^{\ell}_{\tt int}$, $C^{\ell}_{\tt mix}$ and $C^{\ell}_{\tt \gamma \gamma}$ are discussed in \cite{Ecker:1987hd,Ecker:1987qi}. Constraint on $C^{(1)}_{lq}/\Lambda^2$ and $C^{(3)}_{lq}/\Lambda^2$ couplings provide prediction on this observable.

	\item \underline{$\mathcal{B}$($D^0 \to \mu \mu$)}: Branching ratio of $D^0 \to \mu \mu$ decay is noted as \cite{Golowich:2009ii},
	\beq
	{\rm BR}(D^0 \to \mu \mu)= \frac{M_D}{8 \pi \Gamma_D \Lambda^4}\sqrt{1-\frac{4 m_{\mu}^2}{M_D^2}}\left[\left(1-\frac{4 m_{\mu}^2}{M_D^2}\right)|A|^2+|B|^2\right],
	\eeq 
	with, 
	\begin{align}
	\begin{split}
	A=&\frac{G_F f_D M_D^2}{2 m_c}C^{(1)31}_{l e q u},\\
	B=&\frac{G_F f_D m_{\mu}}{2 m_c}(C^{(1)31}_{l q}+C^{(3)31}_{l q}-C^{31}_{l u}+C^{31}_{q e})),\\
	\end{split}
	\label{eq:ab}
	\end{align}
	where $m_c$ is the mass of charm quark, $M_D$ and $f_D$ are the mass and decay constant of $D^0$, respectively. Tensor mediated operator does not contribute to this observable as $\left< \ell^+ \ell^- | \mathcal{O}^{(3)}_{l e q u} | D^0 \right>=0$. Constraints on operators written in Eq.~\eqref{eq:ab} provide 
	prediction on this observable.
	
		\begin{figure}[htb!]
		$$
		\includegraphics[height=4.5cm,width=7cm]{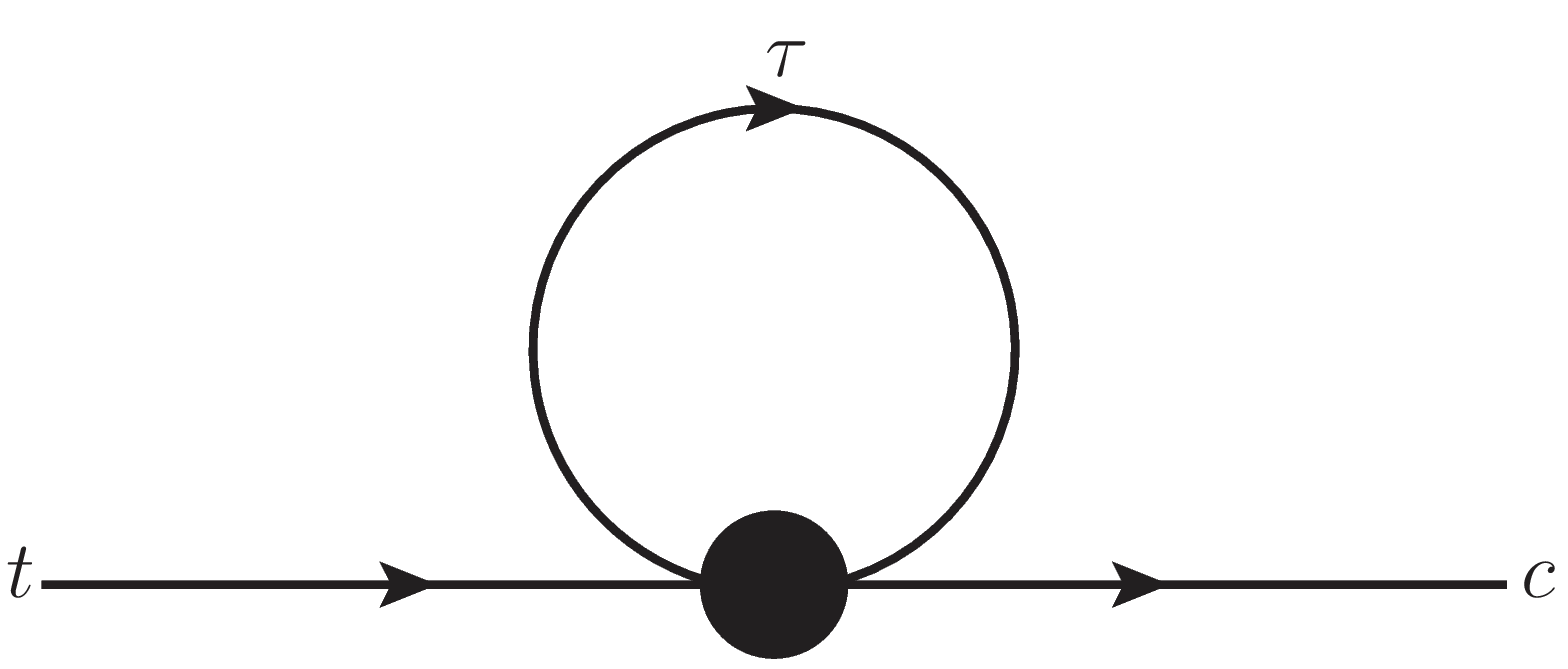}
		$$
		\caption{$\tau$ mediated 1-loop scalar mediated operator $\mathcal{O}_{l e q u}^{(1)}$ contribution to $t \to c \gamma$ decay where a photon is radiated from $t$, $c$ leg, and $\tau$ loop. }
		\label{fig:1loop.tcgamma}
	\end{figure}
	
	\item \underline{$\mathcal{B}$($t \to c \gamma$)}: Scalar mediated dimension-six effective coupling provides a prediction on BR($t \to c \gamma$) through $\tau$ mediated 
	1-loop diagram as shown in figure~\ref{fig:1loop.tcgamma}. Vector and tensor mediated operators do not contribute to this observable as 
	$\left< c \gamma | {\rm any~vector~operator} | t \right>=0$ and $\left< c \gamma | \mathcal{O}^{(3)}_{l e q u} | t \right>=0$. The expression of 
	$\mathcal{B}$($t \to c \gamma$) is given by,
	\beq
	\mathcal{B}(t \to c \gamma)=\frac{8 e_0^2m_{\tau}^6}{9 \Gamma^t(m_t^2-m_c^2)}\left(\frac{C^{(1)32}_{l e q u}}{\Lambda^2}\right)^2\left(1+2{\rm Log}\left[\frac{\Lambda}{m_{\tau}}\right]\right)^2  (m_c^4+7 m_c^2 m_t^2 + m_t^4).
	\eeq
\end{itemize}
Prediction on different observables as noted in Table~\ref{tab:br.prdctn} suggests that all the experimental predictions on dimension-six effective 
couplings are consistent with existing upper bounds.

\begin{table}[htb!]
	{\renewcommand{\arraystretch}{1.3}%
		\centering
		\scalebox{0.9}{
			\begin{tabular}{ |c|c|c|c|c| } 
				\hline
				\multicolumn{1}{|c}{Dimension-six }&
				\multicolumn{1}{|c}{Observables}&
				\multicolumn{2}{|c|}{Branching ratio}\\
				\cline{3-4}
				couplings&  & Prediction & Upper bound  \\ 
				\hline
				&$\mathcal{B}$($K_L \to \pi^0 e^+e^-$)&$7.38 \times 10^{-11}$& $2.8 \times 10^{-10}$ \cite{KTeV:2003sls}\\
				&$\mathcal{B}$($K_L \to \pi^0 \mu^+\mu^-$)&$8.43 \times 10^{-11}$& $3.8 \times 10^{-10}$ \cite{KTEV:2000ngj}\\
				$C^{(1)32}_{l q}/\Lambda^2$&$\mathcal{B}$($D^0 \to \mu \mu$)&$4.63 \times 10^{-23}$& $3.1 \times 10^{-9}$ \cite{LHCb:2022jaa}\\
				&$\mathcal{B}$($ t \to c e e$)&$1.28 \times 10^{-11}$&$2.1 \times 10^{-4}$ \cite{Altmannshofer:2023bfk}\\
				&$\mathcal{B}$($ t \to c \mu \mu$)&$1.30 \times 10^{-11}$&$1.5 \times 10^{-4}$ \cite{Altmannshofer:2023bfk}\\
				\hline
				&$\mathcal{B}$($K_L \to \pi^0 e^+e^-$)&$7.42 \times 10^{-11}$&$2.8 \times 10^{-10}$ \cite{KTeV:2003sls}\\
				&$\mathcal{B}$($K_L \to \pi^0 \mu^+\mu^-$)&$8.46 \times 10^{-11}$&$3.8 \times 10^{-10}$ \cite{KTEV:2000ngj}\\
				$C^{(3)32}_{l q}/\Lambda^2$&$\mathcal{B}$($D^0 \to \mu \mu$)&$4.63 \times 10^{-21}$& $3.1 \times 10^{-9}$ \cite{LHCb:2022jaa}\\
				&$\mathcal{B}$($ t \to c e e$)&$1.31 \times 10^{-11}$&$2.1 \times 10^{-4}$ \cite{Altmannshofer:2023bfk}\\
				&$\mathcal{B}$($ t \to c \mu \mu$)&$1.33 \times 10^{-11}$&$1.5 \times 10^{-4}$ \cite{Altmannshofer:2023bfk}\\
				\hline
				&$\mathcal{B}$($D^0 \to \mu \mu$)& $4.04 \times 10^{-21}$&$3.1 \times 10^{-9}$ \cite{LHCb:2022jaa}\\
				$C^{32}_{qe}/\Lambda^2$&$\mathcal{B}$($ t \to c e e$)&$2.79 \times 10^{-12}$&$2.1 \times 10^{-4}$ \cite{Altmannshofer:2023bfk}\\
				&$\mathcal{B}$($ t \to c \mu \mu$)&$2.82 \times 10^{-12}$&$1.5 \times 10^{-4}$ \cite{Altmannshofer:2023bfk}\\
				\hline
				&$\mathcal{B}$($ t \to c e e$)&$4.60 \times 10^{-13}$&$2.1 \times 10^{-4}$ \cite{Altmannshofer:2023bfk}\\
				$C^{(1)32}_{lequ}/\Lambda^2$&$\mathcal{B}$($ t \to c \mu \mu$)&$4.66 \times 10^{-13}$&$1.5 \times 10^{-4}$ \cite{Altmannshofer:2023bfk}\\
				&$\mathcal{B}$($t \to c \gamma$) &$8.45\times 10^{-17}$& $1.8 \times 10^{-4}$\cite{ATLAS:2019mke}\\
				\hline
				\multirow{2}{*}{$C^{(3)32}_{lequ}/\Lambda^2$}&$\mathcal{B}$($ t \to c e e$)&$2.84 \times 10^{-13}$&$2.1 \times 10^{-4}$ \cite{Altmannshofer:2023bfk}\\
				&$\mathcal{B}$($ t \to c \mu \mu$)&$2.86 \times 10^{-13}$&$1.5 \times 10^{-4}$ \cite{Altmannshofer:2023bfk}\\
				\hline
		\end{tabular}}
		\caption{Prediction of different observables using more conservative flavor contraints WCs from the simultaneous fit using all possible the flavor violating contributions and existing upper bound.}
		\label{tab:br.prdctn}}
\end{table}
\section{Collider analysis}
\label{sec:col}
In this section, we analyze the possibility of probing the NP at collider in terms of the effective operators as in Eq.~\eqref{eq:ffops}. 
For that we focus on those most stringent limits of the dimension-six effective operators stemmed from the flavor observables. If the most stringent limits 
on the NP couplings can be probed for a given $\sqrt{s}$ then the others can be probed at a smaller $\sqrt{s}$. From Section~\ref{sec:flav.cons}, 
we see that the flavor constraint on $C_{lequ}^{(3)32}/\Lambda^{2}$, the tensor coupling is one of the most stringent ones which are on the order of 
$\sim 10^{-11}$ GeV$^{-2}$ followed by a conservative estimate of  $C_{lequ}^{(1)32}/\Lambda^{2}\sim10^{-9}$ GeV$^{-2}$ and
$C^{(1)32}_{lq}/\Lambda^2\sim 10^{-9}$ GeV$^{-2}$. Therefore, in our subsequent analysis, we will focus on these three WCs mostly. 
From Eq.~\eqref{eq:xsec}, it is clear that the signal cross-section exhibits a linear growth with the square of the CM energy, conversely, 
SM background processes are expected to decrease as $\sqrt{s}$ increases. 
Therefore, at high $\sqrt{s}$, a muon collider should have the capability to detect a such NP scenarios where the CM energy of the machine is 
expected to go upto 30 TeV. For our analysis, we consider $\sqrt{s}$ = 10 TeV with an integrated luminosity ($\mathfrak{L}_{\tt int}$) of  1 $\rm{ab^{-1}}$. 
This choice is within the reach of the future muon collider projections \cite{Black:2022cth}. We further note that when we are probing 
$C/\Lambda^{2}\sim 10^{-11}$ GeV$^{-2}$ at $\sqrt{s}$ = 10 TeV, we assume, $\Lambda \gtrsim 10$ TeV, so that the WC 
$C\gtrsim 10^{-3}$, to keep the effective theory framework validated. As mentioned before, the constraint on scalar operator $C_{lequ}^{(1)32}/\Lambda^{2}$ 
from $3 \to 2$ transition observable is less stringent, therefore, it can be probed at the electron-positron colliders at much lower CM energy, which we discuss in Section~\ref{sec:scalar}.

\subsection{$tc ~(\bar{t}c+t\bar{c})$ production cross-section}
The differential $t\bar{c}(\bar{t}c)$ production cross-section at $\mu^+\mu^-$ collider, governed by the effective four-Fermi
contact interaction as shown in figure~\ref{fig:tcprod}, in terms of vector, scalar and tensor couplings is given by\footnote{In the expression of differential cross-section, muon mass is neglected.},
\begin{align}
\nonumber
\frac{d\sigma}{d \cos \theta} = & \frac{3\mathcal{F}}{8} \bigg\{2(\mathcal{V}^2_{RR}+\mathcal{V}^2_{LL})\left( 1+(1+\beta)\cos \theta  +\beta \cos^2 \theta \right) + 2(\mathcal{V}^2_{RL}+\mathcal{V}^2_{LR}) \left( 1-(1+\beta)\cos \theta  +\beta \cos^2 \theta \right)  \\
&+ \mathcal{S}^2_{RR} (1+\beta) -4\mathcal{S}_{RR}\mathcal{T}_{RR}(1+\beta) \cos \theta +16 \mathcal{T}^2_{RR} (1-\beta + 2\beta \cos^2 \cos \theta) \bigg\},
\label{eq:diffx}
\end{align}
where $\mathcal{F}=\frac{s}{\Lambda^4}\frac{\beta^2}{4\pi (1+\beta^3)}$ and $\beta=\frac{s-m^2_t}{s+m^2_t}$. 
Therefore, total production cross-section is,
\beq
\sigma_{\tt prod}=\mathcal{F}\left[8\mathcal{T}^2_{RR}(3-\beta)+\frac{3}{2}\mathcal{S}^2_{RR}+(\mathcal{V}_{LL}^2+\mathcal{V}_{RR}^2+\mathcal{V}_{LR}^2+\mathcal{V}_{RL}^2)(3+\beta)\right].
\label{eq:xsec}
\eeq

\begin{figure}[htb!]
	$$
	\includegraphics[height=6cm,width=7.5cm]{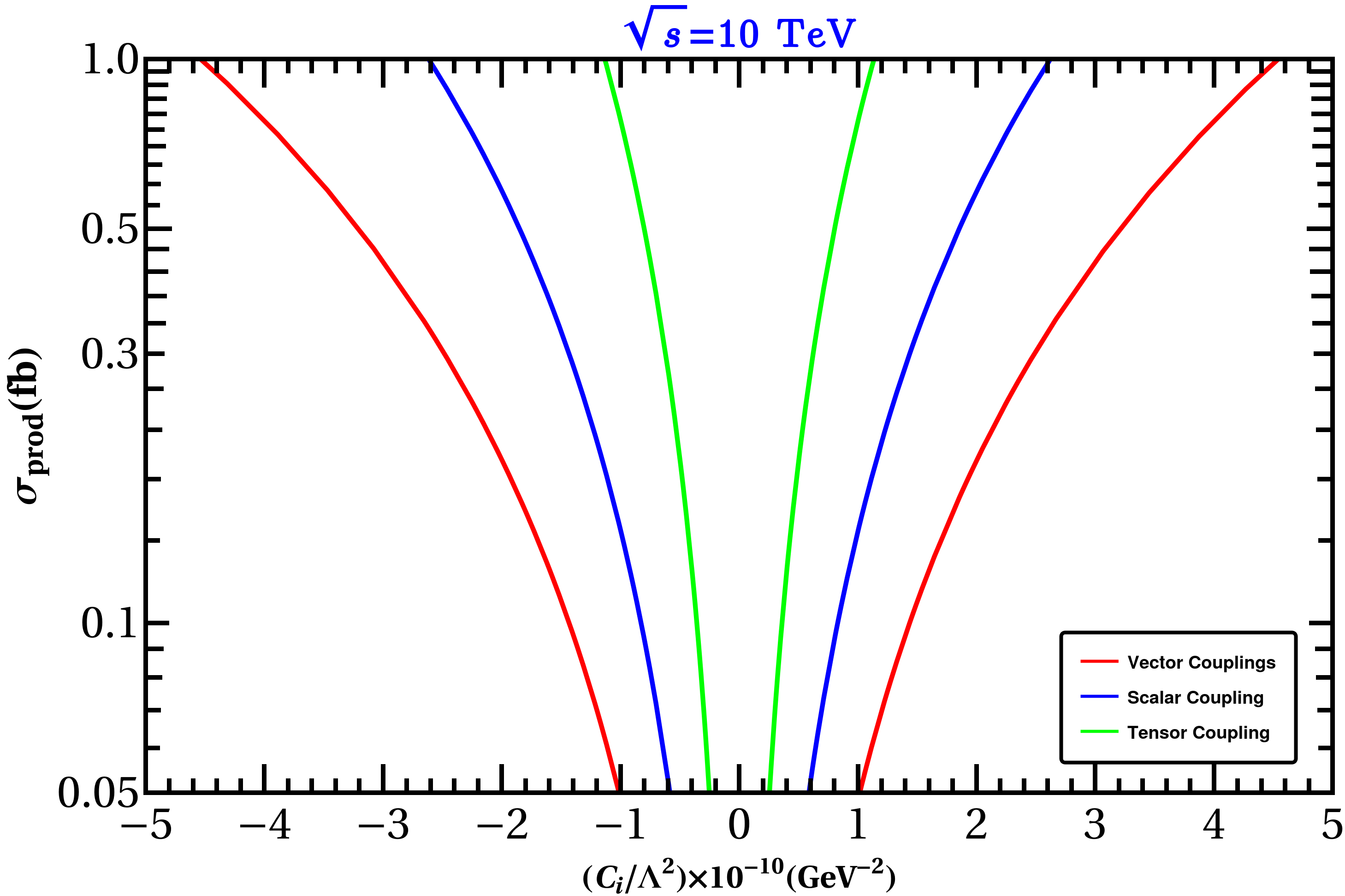}~~~
	\includegraphics[height=6cm,width=7.5cm]{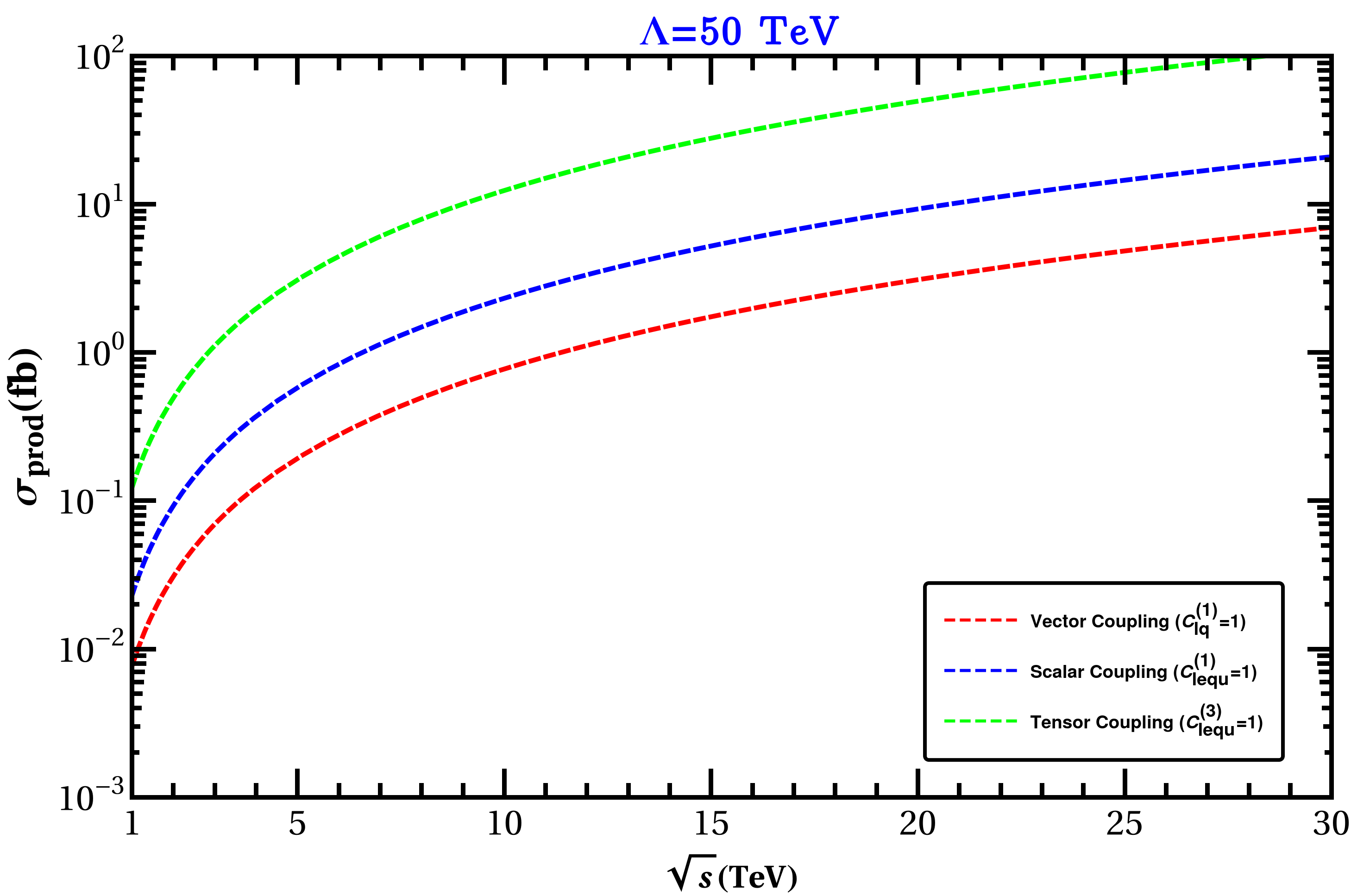}
	$$
	\caption{Variation of total cross-section of $t c$ production at $\mu^+ \mu^-$ collider with different choices of $C/\Lambda^2$ (GeV$^{-2}$) (left) 
		and CM energy ($\sqrt{s}$) for $\Lambda=50$ TeV (right).}
	\label{fig:sigvar}
\end{figure}

\noindent
The variation of the signal cross-section ($\sigma_{\tt prod}$) for the vector, scalar and tensor four Fermi operators used in this model with CM energy ($\sqrt{s}$) 
is shown in the right side of the figure~\ref{fig:sigvar}. Here, $\Lambda=50$ TeV and WCs ($C=1$) are kept constants. 
From Eq.~\eqref{eq:diffx} we see that the total cross-section ($\sigma_{\tt prod}$) is proportional to 
$s/\Lambda^4$, therefore, larger CM energy yields larger cross-section, provided we are in the effective theory limit, i.e. $\sqrt{s}<\Lambda$. 
We also note from figure \ref{fig:sigvar}, that the production cross-section is largest for the tensor coupling and the lowest for the vector coupling. 
For vector coupling, the cross-section $\sim 0.1$ fb at $\sqrt{s}=5$ TeV and that is down to $0.001$ fb at $\sqrt{s}=1$ TeV for 
$C/\Lambda^2\sim 10^{-10}~\rm GeV^{-2}$, making it a necessity to probe these couplings at $\sqrt{s}=10$ TeV.
The variation is shown in one operator scenario, that implies $\sigma_{\tt prod} \propto C^2_i/\Lambda^4$. 
Therefore the variation of total cross-section with $C/\Lambda^2$ is symmetric as depicted in the left side of figure~\ref{fig:tcprod}. 
Apart from the four-Fermi couplings, $Ztc$ couplings ($a^Z_L$ and $a^Z_R$) via $\mathcal{O}^{(1)}_{\phi q}$, $\mathcal{O}^{(3)}_{\phi q}$ 
and $\mathcal{O}_{\phi u}$ also contribute to $tc$ production by interfering with the four-Fermi vector couplings 
$\mathcal{V}_{ij}$. The contribution of $Ztc$ couplings can be incorporated to the total cross-section by redefining, 
\beq
\mathcal{V}_{ij} \to \mathcal{V}_{ij} + 4 c_i^Z a_j^Z \frac{m_W m_Z}{s-m_Z^2}~~~{ \rm{with}} ~~~a^Z_L=\frac{1}{4 c_w}(C_{\phi q}^{(1)}-C_{\phi q}^{(3)})\, , \quad a^Z_L=\frac{1}{4 c_w} C_{\phi u},
\eeq
where, $i,j=L,R$, and $c^Z_L=-1/2+s_{w}^2$ and $c_R^Z=s_{w}^2$ are the couplings to a left or right handed electron, respectively. 
On contrary to four-Fermi couplings, the cross-section via $Z$ mediation drops as $\sim 1/s$ due to the s-channel mediation. In the left side of 
figure~\ref{fig:sigvar}, we show the variation of $\sigma_{\tt prod}$ with varying $C/\Lambda^2$ in GeV$^{-2}$. From the figure, we again see that the tensor 
mediated four-Fermi coupling ($\mathcal{T}_{RR}$) provides the maximum contribution to the $tc$ production and vector mediated production provides the least. In one operator scenario, all the vector operator contribute equally to the total cross-section. 
\begin{table}[h]
	\centering
	\scalebox{1.2}{
		\begin{tabular}{ |c |c | c |}
			\hline
			Couplings ($\rm{GeV^{-2}}$) & Cross-section ($\sigma_{\tt prod}$) ($\rm{fb}$) \\ 
			\hline
			$C^{(1)32}_{lq}/\Lambda^2$ & $4.8 \times 10^{-2}$  \\  
			$C^{(1)32}_{l equ}/\Lambda^2$ & $0.15$  \\ 
			$C^{(3)32}_{l equ}/\Lambda^2$ & $0.77$ \\ 
			$C^{(1)32}_{\phi q}/\Lambda^2$ & $1.71 \times 10^{-10}$ \\ 
			\hline 
	\end{tabular}}
	\caption{$t\bar{c}(t\bar{c})$ production cross-section ($\sigma_{\tt prod}$) at 
		muon collider for vector ($C^{(1)32}_{lq}/\Lambda^2$), scalar ($C^{(1)32}_{l equ}/\Lambda^2$), 
		tensor ($C^{(3)32}_{l equ}/\Lambda^2$), and $Ztc$ ($C^{(1)32}_{\phi q}/\Lambda^2$) effective couplings. We consider $\sqrt{s}$ = 10 TeV and 
		$C/\Lambda^2=10^{-10} ~\rm GeV^{-2}$ for all the operators.}
	\label{tab:xsec}  
\end{table}
If we assume $Ztc$ effective couplings to be of the same order to that of vector coupling,  $C/\Lambda^2\sim 10^{-9}~ \rm GeV^{-2}$, 
then the contribution from $Ztc$ is way milder than the four-Fermi operator contribution, see Table~\ref{tab:xsec}, where we see that the production cross-section ($\sigma_{\tt prod}$) for vector coupling surpasses that of $Ztc$ coupling by an order of $10^8$ in fb at 10 TeV muon collider. Therefore in the following analysis we neglect the contribution of $Ztc$ and study the four-Fermi operators. We refrain from studying dimension-eight contributions, which in principle could contribute at the same order as our considered dimension-six contributions.

\subsection{Signal and background processes}

In high-energy muon colliders, collision events take place at a CM energy surpassing that achieved by the present LHC (parton level) or potential future electron-positron colliders. Consequently, it is anticipated that jets which are closely clustered together in these collisions, will exhibit 
collimation, effectively coalescing into a singular, consolidated jet. A notable illustration of this behavior can be observed in the jets stemming from the 
hadronic decay processes of top quarks or W/Z bosons, where they converge to form a single ``top" or ``W/Z'' jet \cite{Duan:2022nuy}. Since, the signal process is 
$\mu^{-} \mu^{+} \xrightarrow{} tc$ (by $tc$ production we always mean $\bar{t}c+t\bar{c}$ production), for this analysis, we consider a di-jet signal with no leptons. The corresponding SM background processes are $\mu^{-} \mu^{+} \xrightarrow{} t \overline{t}$, $\mu^{-} \mu^{+} \xrightarrow{} q \overline{q}$, $\mu^{-} \mu^{+} \xrightarrow{} c \overline{c}$, $\mu^{-} \mu^{+} \xrightarrow{} b \overline{b}$, $\mu^{-} \mu^{+} \xrightarrow{} W^{+} W^{-}$ and $\mu^{-} \mu^{+} \xrightarrow{} ZZ$. 

\subsubsection{Event simulation} 
\label{sec:es}
The signal and background events are generated in \texttt{MadGraph5} \cite{Alwall:2014hca}. For the EFT signal process, the \texttt{UFO} model file is generated using \texttt{FeynRules} \cite{Christensen:2008py}. The benchmark points (BPs) we consider for collider analysis are:
\begin{itemize}
	\item BP1 (Vector) :  $C^{(1)32}_{lq}/\Lambda^{2} = -9.96 \times 10^{-10}$ GeV$^{-2}$,
	\item BP2 (Scalar) : $C_{lequ}^{(1)32}/\Lambda^{2} = -9.27 \times 10^{-10}$ GeV$^{-2}$,
	\item BP3 (Tensor) : $C_{lequ}^{(3)32}/\Lambda^{2} = 7.24 \times 10^{-11}$ GeV$^{-2}$.		
\end{itemize}
These BPs are most stringent flavor constrained vector, scalar, and tensor couplings as noted in Table~\ref{tab:cplng.const} and \ref{tab:cplng.const2}. 
Using these numerical inputs, Monte-Carlo (MC) generated events are fed into \texttt{Pythia8} \cite{Sjostrand:2014zea} and \texttt{Delphes3} \cite{deFavereau:2013fsa} for 
parton showering and detector simulation, respectively. For event simulation, following criteria are used:
\begin{itemize}
	\item The lepton selection is done in  \texttt{Delphes3} with criteria $p_{T} >$ 10 GeV. Further, we impose $|\eta| \leq$ 2.5 for electrons and $|\eta| \leq$ 2.4 for muons. For electrons, in the allowed $|\eta|$ region, the detection efficiencies are 0.95 in $|\eta| \leq$ 1.5 region and 0.85 in  1.5 $< |\eta| \leq$ 2.5 region. For muon the detection efficiency is 0.95 throughout the allowed region. The lepton isolation criteria is set $\Delta R >$ 0.5 from another lepton or jet.
	\item The jet clustering has been done in \texttt{Fastjet3} \cite{Cacciari:2011ma} using \texttt{anti-kt} algorithm \cite{Cacciari:2008gp}. 
	The jet radius is taken to be 0.5 with the minimum $p_{T}$ of the jet set as 20 GeV. The jet reconstruction is done using \texttt{Delphes3}.
\end{itemize}

\begin{figure}[htb!]
	\centering
	$$
	\includegraphics[width= 7.5 cm, height= 5.7 cm]{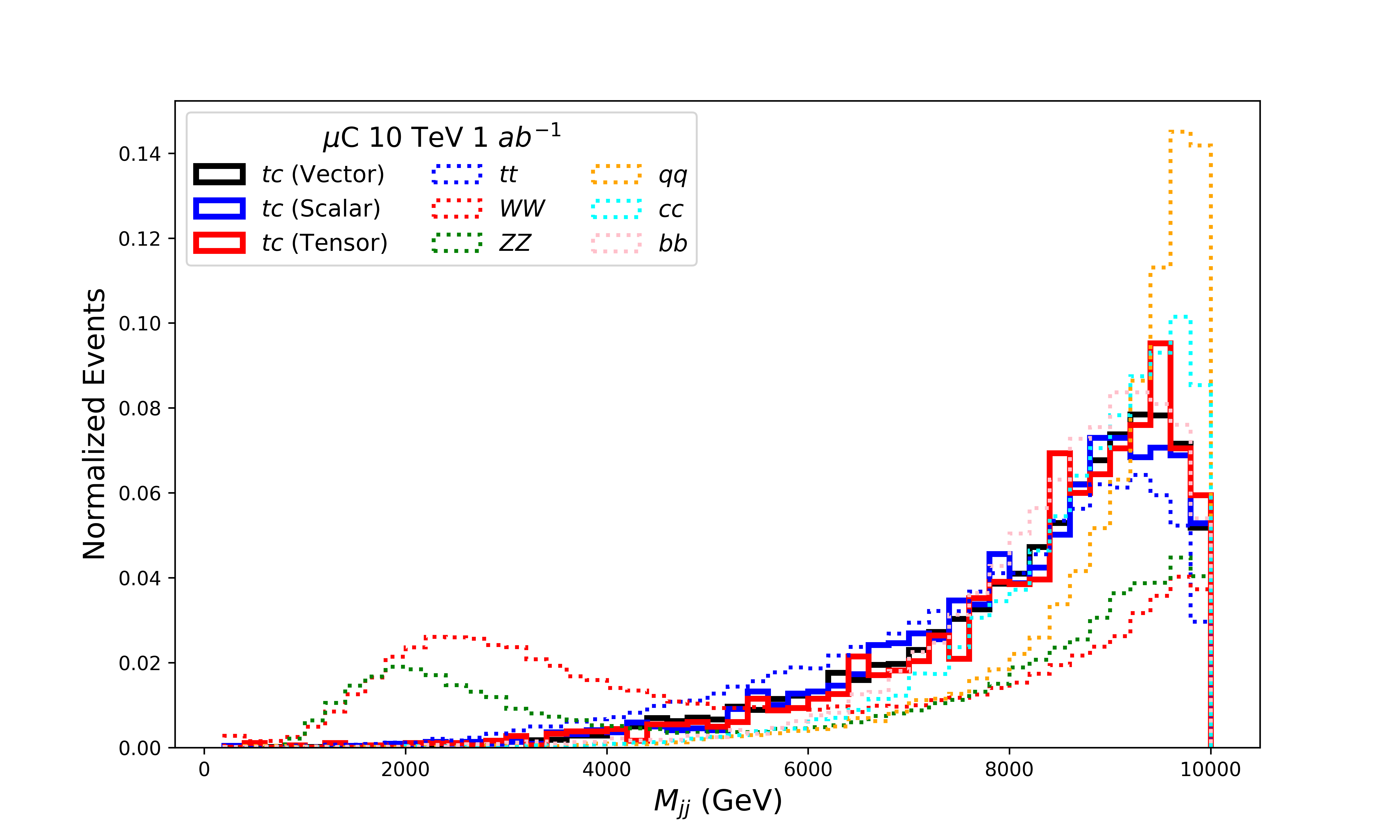}~~
	\includegraphics[width= 7.5 cm, height= 5.7 cm]{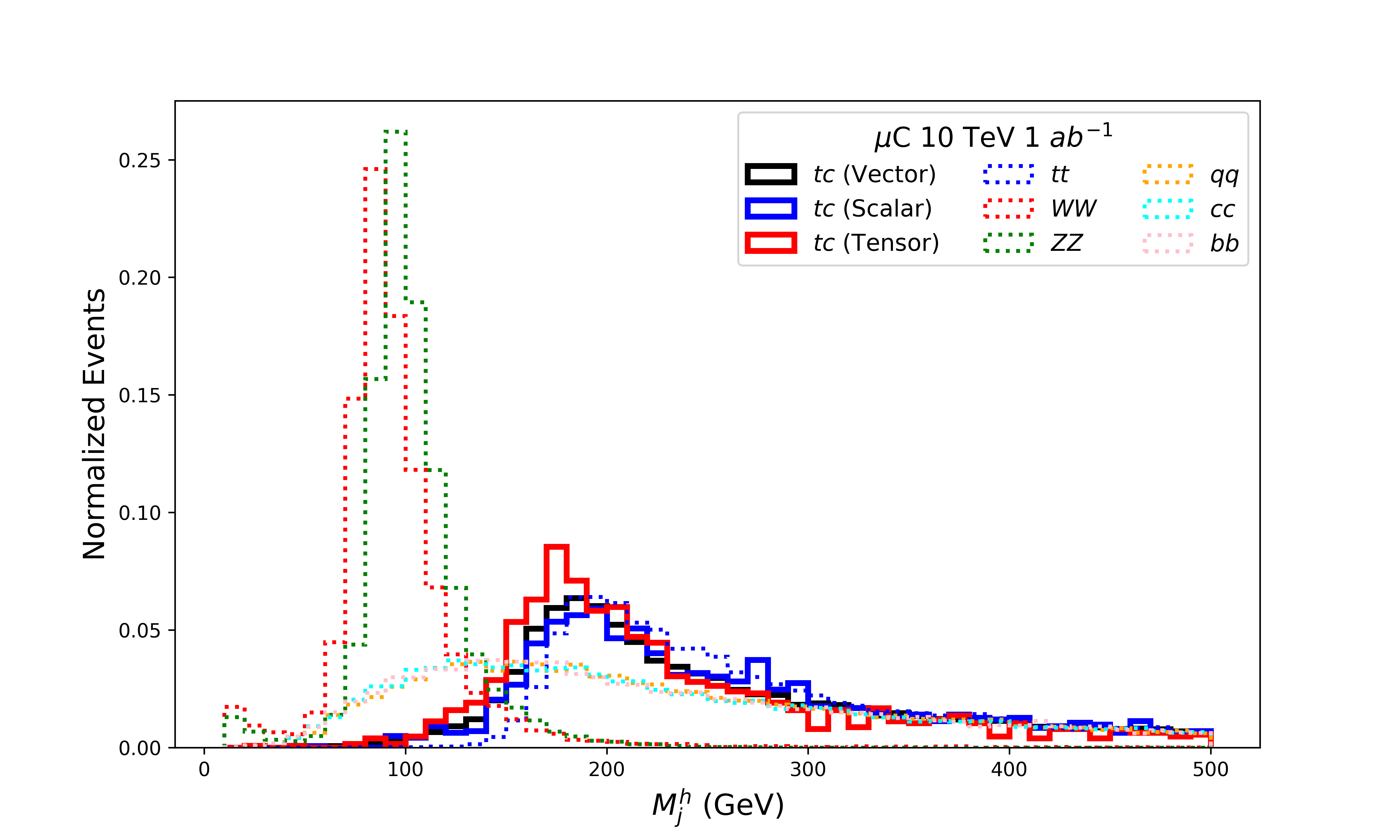}
	$$
	\caption{Normalized event distribution for invariant di-jet mass ($M_{jj}$) (left) and heavy jet mass ($M_j^{h}$) (right) in case of di-jet final state containing signal and background processes at the muon collider with $\sqrt{s}=10$ TeV and $\mathfrak{L}_{\tt int}=$ 1 $\rm{ab}^{-1}$.
		\label{fig:event.dist}}
\end{figure}

\subsection{Cut based Analysis}
For the analysis, we use the following sequential cuts for the signal and background processes:
\begin{itemize}
	\item Cut 1: $N_{j} =$ 2 and $N_{l} =$ 0.
	\item Cut 2: $M_{jj} > $ 8 TeV.
	\item Cut 3: $M_{j}^{h} >$ 160 GeV.
\end{itemize}

Here, $M_{jj}$ is the invariant mass of the di-jet, $M_{j}^{h}$ is the invariant masses of the reconstructed particles of the heavier jet. The major background, in case of our analysis, are the diboson processes. In Cut 1, which is our signal selection cut, we demand only processes with 2 jets. We further remove events with detected leptons. This will remove the detected leptonic and semi-leptonic decay processes for $W^{+}W^{-}$, $ZZ$ as well as $t \overline{t}$. The invariant mass of the di-jet pair is expected to peak near the CM energy of the incoming particles for jets coming from the production level processes. For jets coming from branching or radiation, the invariant mass is expected to peak at a lower value. Imposing Cut 2, we can remove further semileptonic and leptonic processes from di-boson production, where the leptons were not detected. Now, due to the collimation of boosted jets in high energies, as we mention above, we expect multiple jets branching from top or W/Z to appear as a single jet in the detector. The invariant mass of the reconstructed particles of such jets are expected to peak at the mass of their respective sources, i.e. top mass or W/Z mass. In $tc$ signal process, the heavier jet is expected to peak at the top mass, as such Cut 3, will significantly remove $W^{+}W^{-}$ and $ZZ$ backgrounds as evident from the distribution in figure~ \ref{fig:event.dist}. The signal can be further separated from the background by using a charm tagging algorithm on the lighter jet of $tc$ signal process. However, with the current c-tagging efficiency at the LHC \cite{ATLAS:2015prs,CMS:2016knj}, it looks like a far-fetched possibility. This is primarily due to high miss-tagging efficiency of the b-jets as c-jets. After employing sequential cuts as mentioned above, signal significance $S/\sqrt{S+B}$ and efficiency factor\footnote{$\sigma_{s(b)}^{\tt sig}$ is the final state cross-section after final for signal (background) whereas $\sigma_{s(b)}^{\tt prod}$ is the production cross-section for signal (background).} $\epsilon_{s(b)}=\sigma_{s(b)}^{\tt sig}/\sigma_{s(b)}^{\tt prod}$ are noted in Table~\ref{tab:events}. We see that the 
benchmark point with vector couplings have most signal significance, while the tensor one has the least.

\begin{table}[htb!]
	\centering
	\begin{tabular}{|l|l|l|l|l|c|c|}
		\hline
		Process & No Cuts & Cut 1 & Cut 2 & Cut 3 & $\epsilon_{s/b}$ & Significance \\ \hline
		$\mu^{-} \mu^{+} \xrightarrow{} tc$ (BP1) & 6432 & 1339 & 891 & 809 & 0.12 & \\
		$\mu^{-} \mu^{+} \xrightarrow{} tc$ (BP2) & 4189 & 918 & 594 & 547 & 0.13 & \\
		$\mu^{-} \mu^{+} \xrightarrow{} tc$ (BP3) & 140 & 30 & 20 & 18 & 0.13 & BP1: 18.40 \\
		$\mu^{-} \mu^{+} \xrightarrow{} t \overline{t}$ & 1730 & 352 & 191 & 183 & 0.10 & \\
		$\mu^{-} \mu^{+} \xrightarrow{} q \overline{q}$ & 3555 & 746 & 650 & 433 & 0.12 & BP2: 13.38\\
		$\mu^{-} \mu^{+} \xrightarrow{} c \overline{c}$ & 1728 & 363 & 290 & 190 & 0.11 & \\
		$\mu^{-} \mu^{+} \xrightarrow{} b \overline{b}$ & 914 & 194 & 143 & 93 & 0.10 & BP3: 0.53 \\ 
		$\mu^{-} \mu^{+} \xrightarrow{} W^{+} W^{-}$ & 58860 & 22882 & 7258 & 215 & 0.004 &  \\
		$\mu^{-} \mu^{+} \xrightarrow{} ZZ$ & 3284 & 1238 & 454 & 10 & 0.003 & \\ \hline		
	\end{tabular}
	\caption{Cut flow and signal significance for signal at the chosen benchmark points
		and background processes for analysis at muon collider at 10 TeV and  1 $\rm{ab^{-1}}$. 
		Here $\epsilon_{s/b}$ denotes the cut efficiency for signal and background following the final cut (Cut 3).
		\label{tab:events}}
\end{table}

\section{Optimal Observable Technique}
\label{sec:oot}
The optimal observable technique (OOT) is a statistical tool that enables optimal estimation of NP couplings via 
minimizing the covariance matrix, which has been used studied for different cases, but mostly without a non-interfering SM background \cite{Atwood:1991ka,Davier:1992nw,Diehl:1993br,Gunion:1996vv}. In general, a collider observable ($e.g.,$ differential cross-section) contains 
contribution from both the signal and background. If we wish to write the observable at the production level having the cut efficiencies of signal analysis 
embedded, then it can be written as,
\beq
\mathcal{O}=\frac{d\sigma_{\tt theo}}{d\phi} = \epsilon_s \mathcal{O}_{\tt sig} + \epsilon_b \mathcal{O}_{\tt bkg} =\sum_i g_i f_i(\phi) \,.
\label{eq:obs}
\eeq
Here $ g_i $ are the non-linear functions of NP couplings and $f_i$ are the function of phase-space co-ordinate $\phi$ and 
$\epsilon_s$ ($\epsilon_b$) is the signal (background) efficiencies in estimating the signal (and background) after using judicious cuts 
as tabulated in Table~\ref{tab:events}. In this analysis as our focus on $2 \to 2$ processes, the phase space co-ordinate will be 
$\phi=\cos \theta$, where $\theta$ is angle between the outgoing particles with respect to 
the beam axis. However, the choice of $\phi$ can vary depending on the specific process. We importantly note that the 
signal of the processes under consideration are essentially di-jet events (as elaborated in the previous section), stemming from the 
produced particles in the high CM energy of the muon collider, enabling us to use Eq.~\eqref{eq:obs}, without much problem. However, 
for processes where the decay products are not collimated, say for example, at smaller CM energy, the usage of Eq.~\eqref{eq:obs} is limited. For a generic procedure of inclusion of SM background, see \cite{Bhattacharya:2023zln}.

The essential idea is to determine the NP coefficients $g_i$ as precisely as possible. In case of a realistic experimental scenario, 
the event numbers obey Poisson distribution, we can estimate $g_i$ with the application of appropriate weighting functions ($w_i(\phi)$), 
\beq
g_i=\int w_i(\phi) \mathcal{O}(\phi) d\phi.
\eeq
There's one particular choice of $w_i(\phi)$ that optimizes the covariance matrix ($\mathbbm{V}_{ij}$) in a sense that statistical uncertainties in 
$g_i$'s are minimal. In this case, $\mathbbm{V}_{ij}$ is expressed as:

\beq
\mathbbm{V}_{ij} =\frac{ \mathbbm{M}_{ij}^{-1} \sigma_T}{N}= \mathbbm{M}^{-1}_{ij}\mathfrak{L}_{\tt int}^{-1}\, ,~~\mathbbm{M}_{ij} =\int \frac{f_i(\phi) f_j(\phi)}{\mathcal{O}(\phi)}d\phi.
\label{eq:cov.mat}
\eeq
Then, the optimal covariance matrix reads 
\beq
\mathbbm{V}_{ij} =\frac{ \mathbbm{M}_{ij}^{-1} \sigma_T}{N}= \frac{\mathbbm{M}^{-1}_{ij}}{\mathfrak{L}_{\tt int}}\, ,
\label{eq:covmat1}
\eeq
with $\sigma_T=\int \mathcal{O}(\phi) d\phi$ defining the total cross-section and $N$ is total number of events 
($N=\sigma_T \mathfrak{L}_{\tt int}$) including the SM background contributions. 
$\mathfrak{L}_{\tt int}$ indicates the integrated luminosity of the collider over a period.
The $\chi^2$ function, which quantifies the precision of NP couplings, is defined as:
\beq
\chi^2= \sum_{\{i,j\}=1}^{n} (g_i -g_i^0) (g_j -g_j^0) \,  \left( \mathbbm{V}^{-1}\right)_{ij},
\label{eq:chi2}
\eeq
where $g^0$'s are the {\it `seed values'} of NP model inputs. Here we use them as the ones predicted from the flavor constraints. 

\subsection{OOT Sensitivities}
\noindent
Here, we discuss the optimal uncertainties and correlations of different dimension-six effective operators in context of $t\bar{c}(\bar{t}c)$ production. 
As mentioned before, we focus on the probe of most stringent vector, scalar, and tensor mediated couplings that contribute to the $tc$ production 
at the $\mu^+ \mu^-$ colliders. 
\begin{figure}[htb!]
	$$
	\includegraphics[height=5cm,width=5cm]{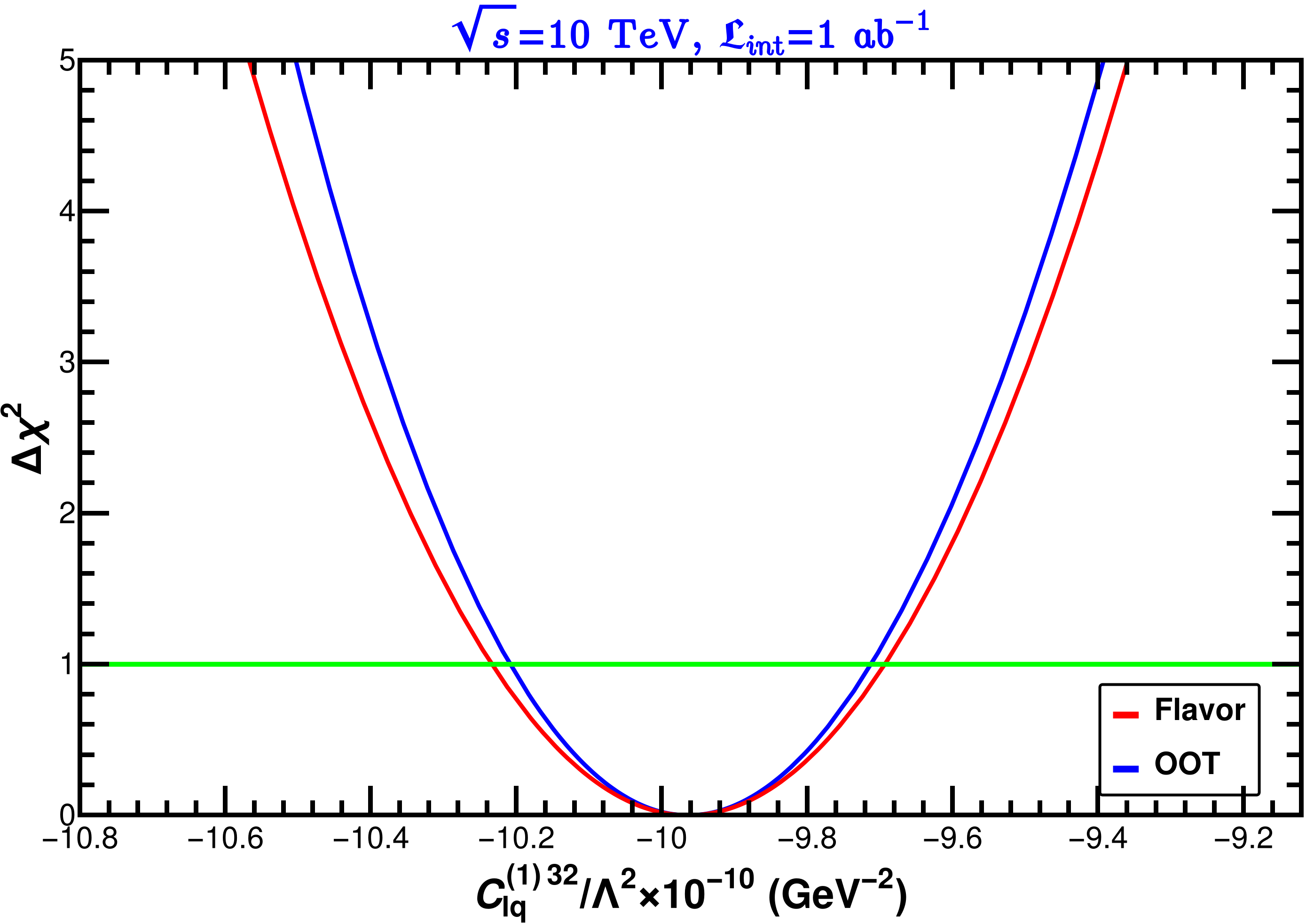}~~
	\includegraphics[height=5cm,width=5cm]{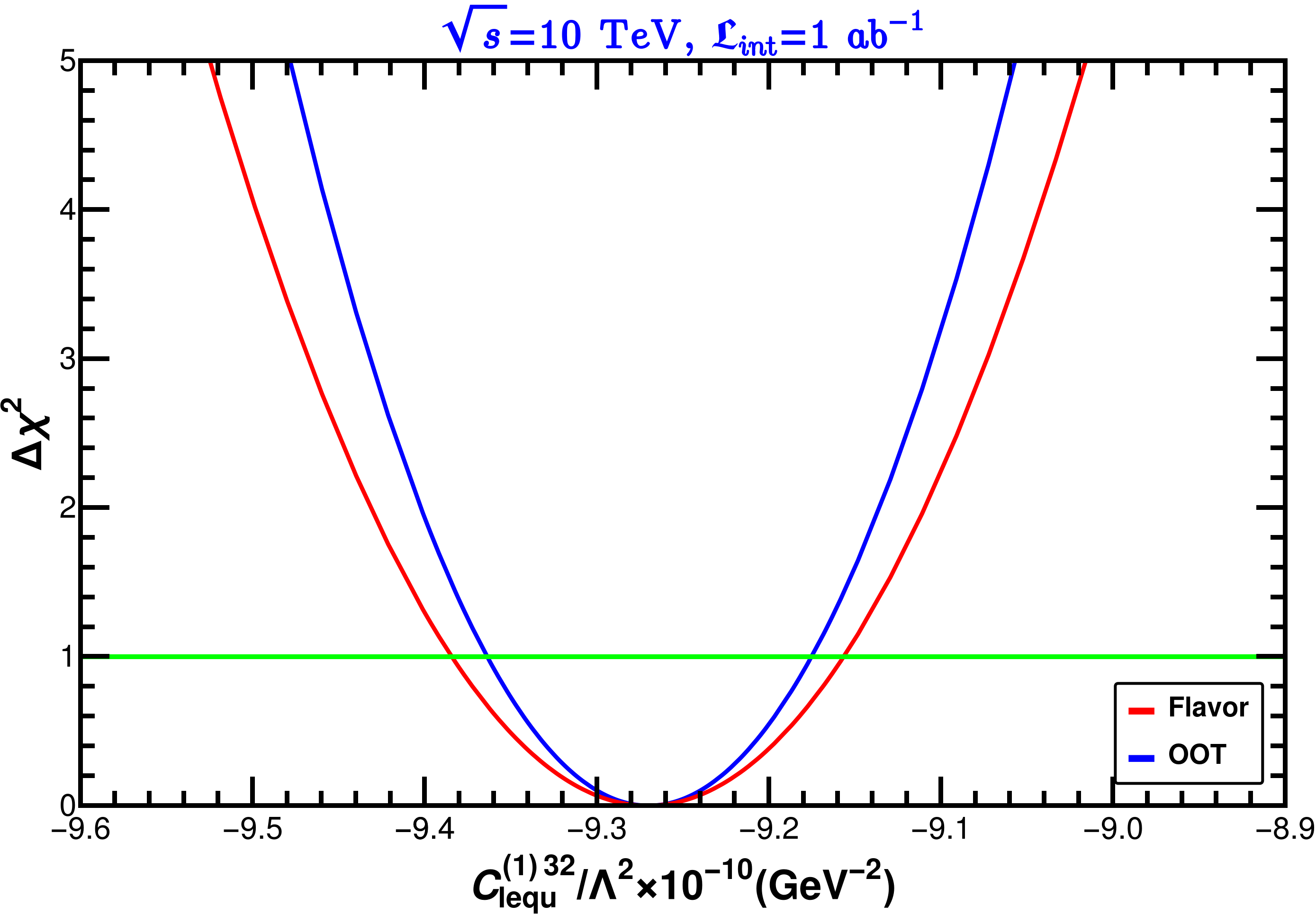}~~
	\includegraphics[height=5cm,width=5cm]{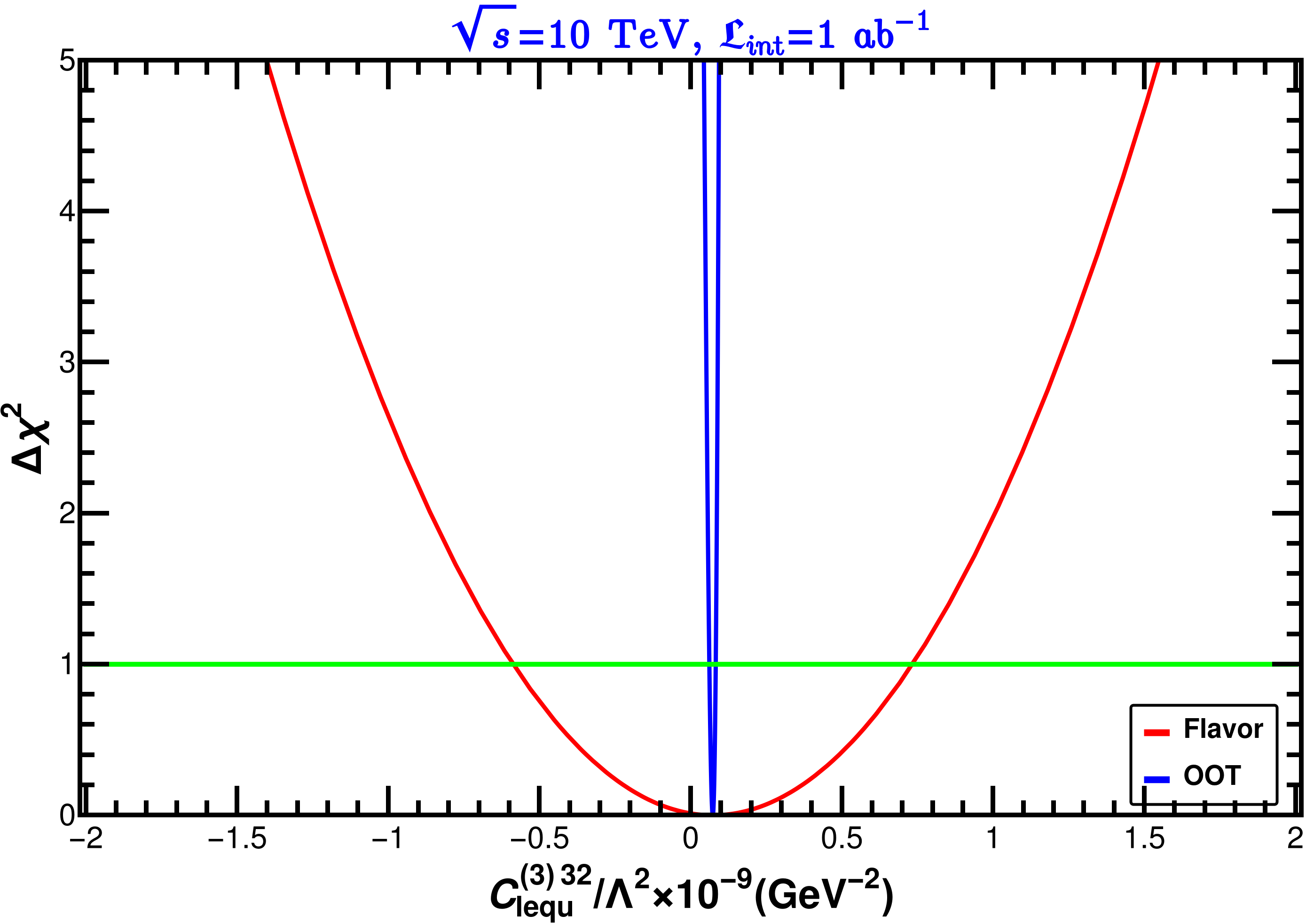}
	$$
	\caption{The $\Delta \chi^2$ variation as a function of dimension-six effective couplings; left: $C^{(1)32}_{lq}/\Lambda^2$ (Vector), middle: $C^{(1)32}_{lequ}/\Lambda^2$ 
	(Scalar),  right: $C^{(3)32}_{lequ}/\Lambda^2$ (Tensor). The units of $C^{ij}/\Lambda^2$ is in $\rm{GeV^{-2}}$. The horizontal green axis denotes $\Delta \chi^2=1$.}
	\label{fig:1sigma.1d}
\end{figure}
The $\chi^2$ variation dimension-six effective operators are shown in the figure~\ref{fig:1sigma.1d}. 
At CM energy $\sqrt{s}$=10 TeV and luminosity $\mathfrak{L}_{\tt int}=1~\rm{ab^{-1}}$, we observe that for all the couplings, optimal uncertainty 
is narrower than flavor uncertainty.  It is intriguing to note that for the tensor-mediated coupling $C_{lequ}^{(3)32}/\Lambda^{2}$, the flavor uncertainty 
is way larger than the optimal collider sensitivity of this coupling at muon collider (see the right most panel in figure~\ref{fig:1sigma.1d}). 
This suggests that the pertinent operator is less-consistent in flavor observable, yet in a collider scenario, it can be measured with larger precision. 
This is because the tensor operator having most stringent flavor bound, has the largest contribution to the $tc$ production compared to scalar and vector operators, 
as already evidenced by figure~\ref{fig:sigvar} and Table~\ref{tab:xsec}. 

\begin{figure}[htb!]
	$$
	\includegraphics[height=6cm,width=7.5cm]{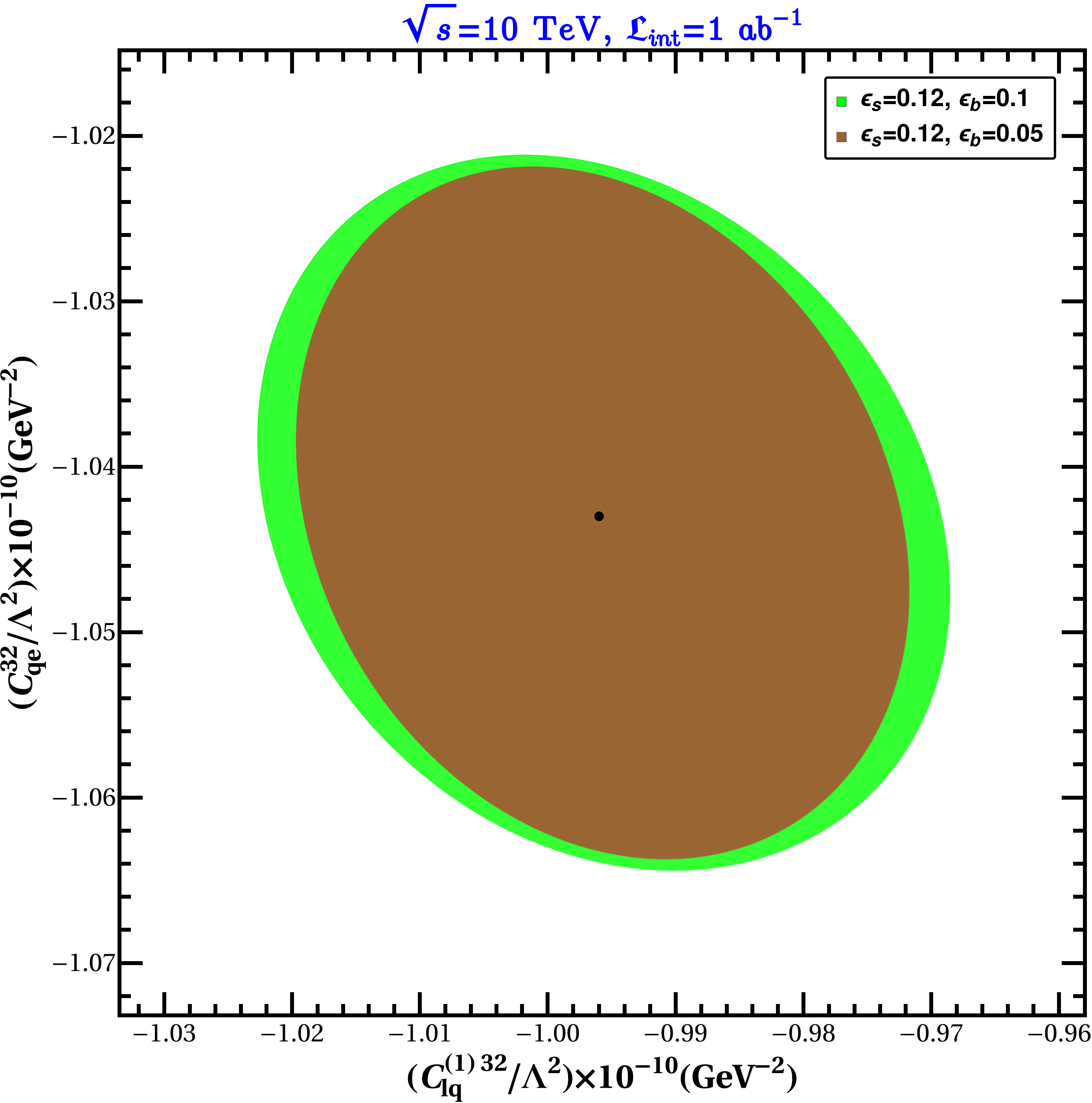}~~
	\includegraphics[height=6cm,width=7.5cm]{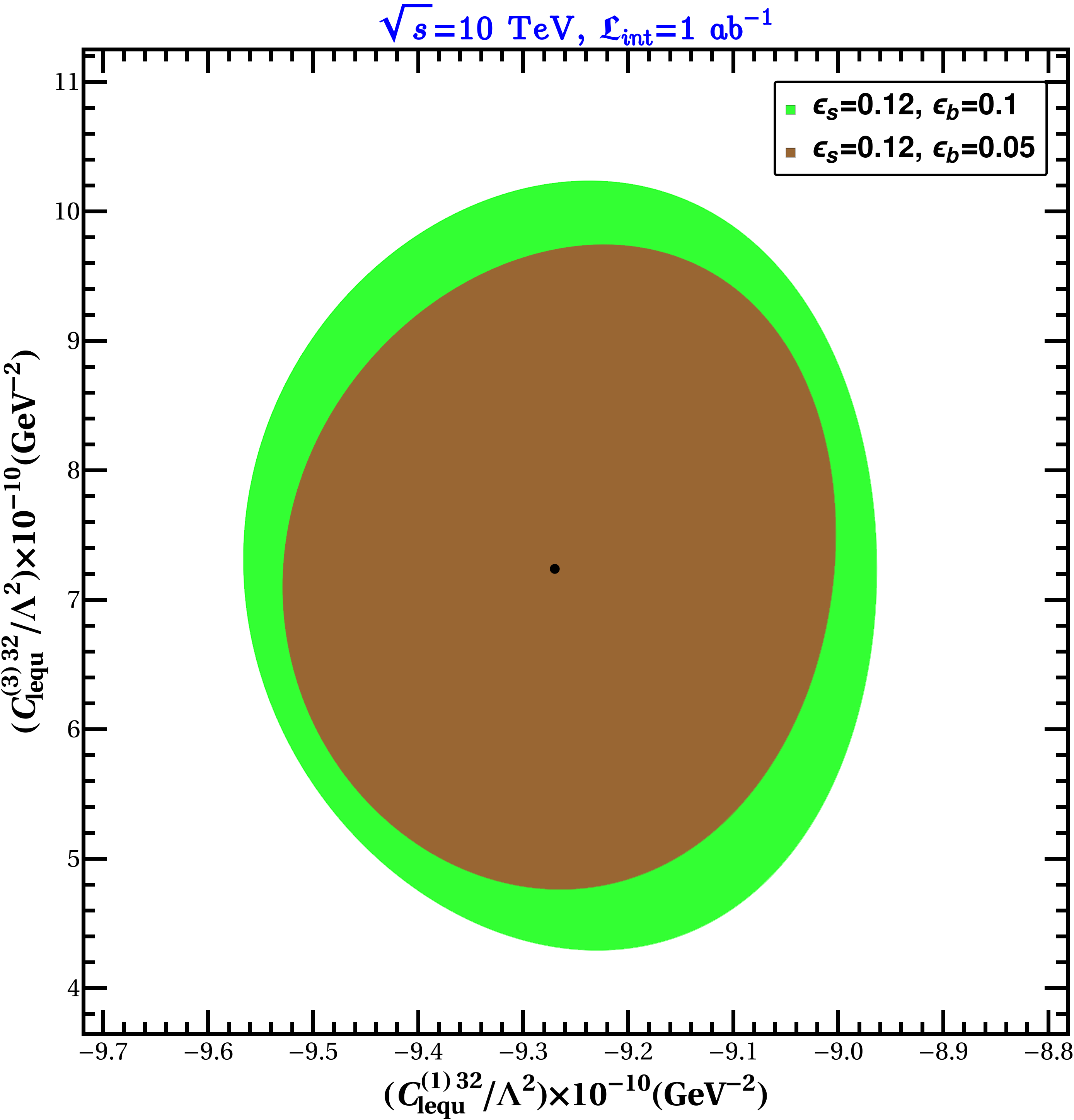}
	$$
	\caption{Optimal 1$\sigma$ allowed parameter space in 2D  plane. Left: $C^{(1)32}_{lq}/\Lambda^2-C^{32}_{qe}/\Lambda^2$ plane; right: $C^{(1)32}_{lequ}/\Lambda^2-C^{(3)32}_{lequ}/\Lambda^2$ plane. Variation of background efficiency ($\epsilon_b$) is also shown in both cases. See inset and headings for details. The flavor inputs are taken from the case-III mentioned in Table~\ref{tab:cases}.}
	\label{fig:epsilon}
\end{figure}

The correlations between the two vector couplings and correlations among scalar and tensor couplings for different background efficiencies 
($\epsilon_b=\{0.1,0.05\}$) are shown in figure~\ref{fig:epsilon}. The benchmark cases with seed values are listed in Table~\ref{tab:cases}.
It is clear that the presence of background via $\epsilon_b$ plays a crucial role in determining the optimal uncertainty of NP couplings, the lesser 
the contamination, the better the precision. Moreover, the optimal uncertainty of NP couplings depends 
on the relative NP signal and non-interfering SM background contribution to the final state. Contribution to the final state from two vectors
operators is greater than the scalar and tensor operators together by a factor of 25 (Table~\ref{tab:events}), due to the choice of the 
seed values of the benchmark points. Therefore, the relative NP signal is larger than the background contribution for the vector operators than the 
scalar and tensor ones. That is why the vector operator correlation is less affected 
by the change in background contamination as shown in the figure~\ref{fig:epsilon}. 

\begin{figure}[htb!]
	$$
	\includegraphics[height=6cm,width=7.5cm]{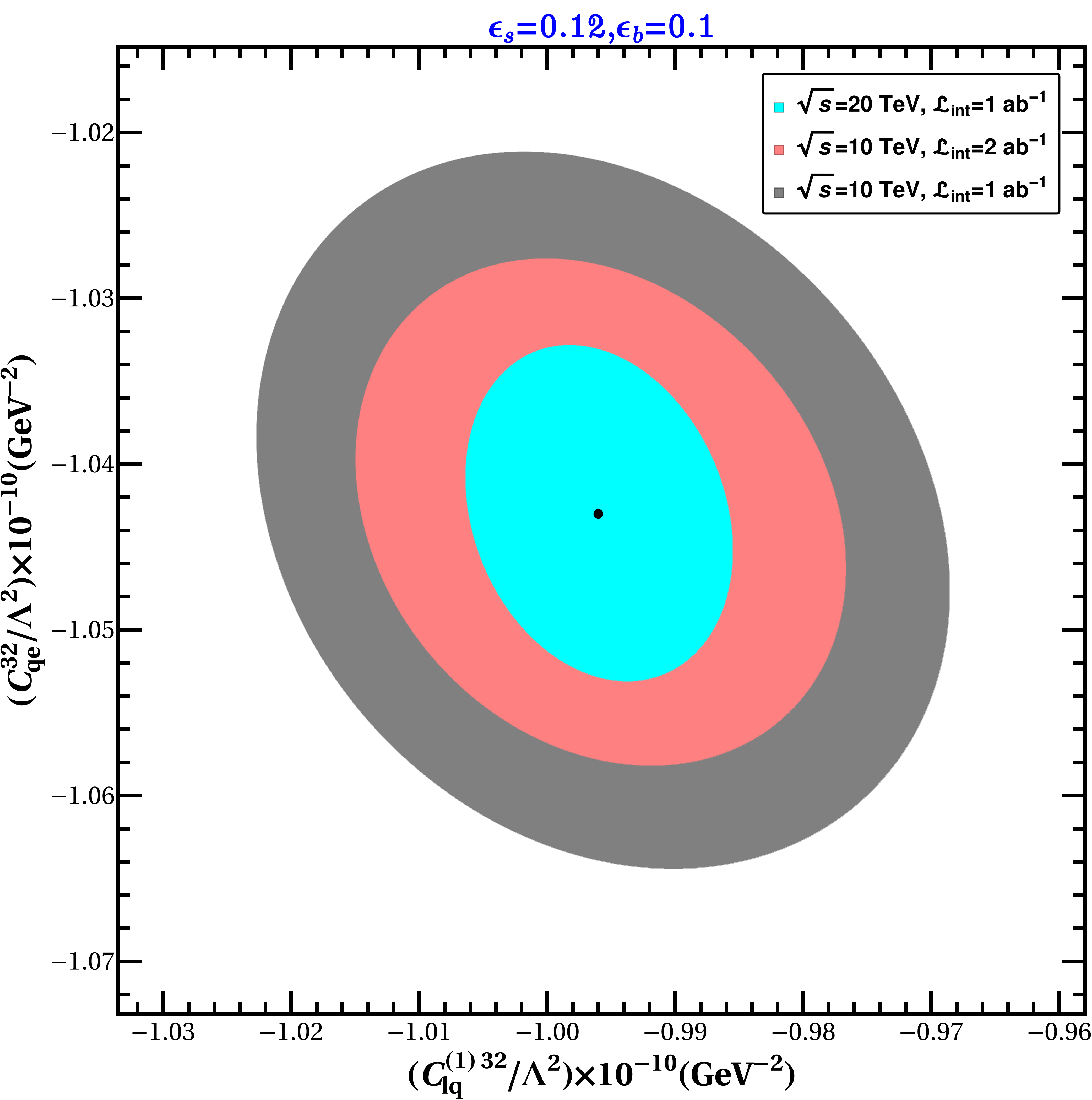}~~
	\includegraphics[height=6cm,width=7.5cm]{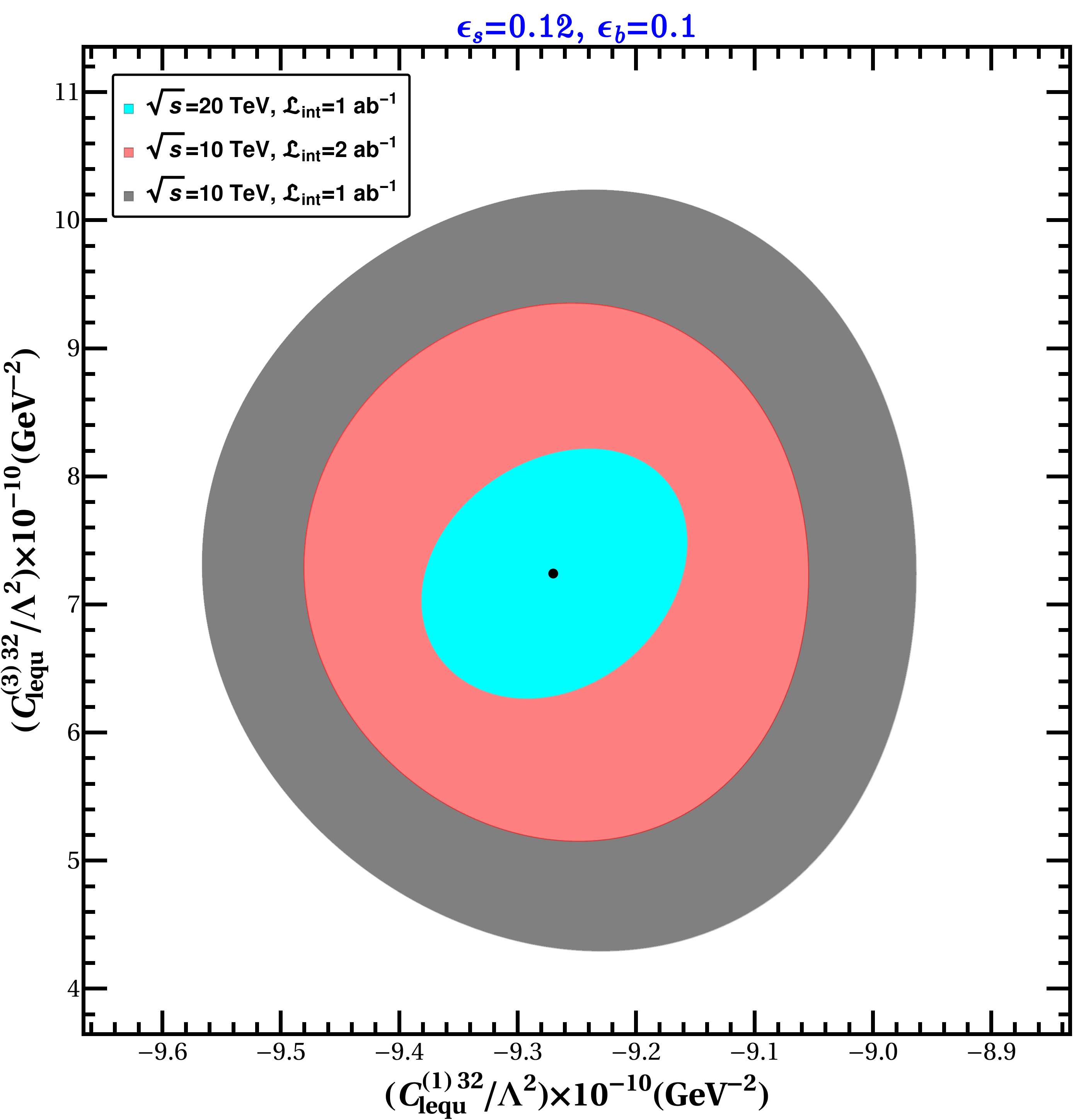}
	$$
	\caption{Comparison of optimal 1$\sigma$ surfaces for CM energy ($\sqrt{s}$) and luminosity ($\mathfrak{L}_{\tt int}$); Left: $C^{(1)32}_{lq}/\Lambda^2-C^{32}_{qe}/\Lambda^2$ plane; right: $C^{(1)32}_{lequ}/\Lambda^2-C^{(3)32}_{lequ}/\Lambda^2$ plane. The flavor inputs are taken from the case-III mentioned in Table~\ref{tab:cases}.}
	\label{fig:cm.lum}
\end{figure}

In  figure~\ref{fig:cm.lum} we show the variation of 1$\sigma$ regions for different CM energies and integrated luminosities. In our scenario, the increase 
of CM energy ($\sqrt{s}$) is more effective than the increase of integrated luminosity. If we increase the CM energy twice, then the signal cross-section is 
increased by a factor of 4 which in turn suggest four times enhancement in signal events. However, we must remember that when we enhance $\sqrt{s}$, 
we choose $\Lambda>\sqrt{s}$ to be consistent with EFT framework, reducing the WC appropriately to keep $C/\Lambda^2$ is same ballpark.
Larger CM energy also reduces the non-interfering SM backgrounds. However, if we double the luminosity, both the signal and background events will 
enhance twice, suggesting larger CM energy helps reducing optimal uncertainty for EFT frameworks as considered here.

\begin{table}[h]
	\centering
	\scalebox{1.1}{
		\begin{tabular}{ |c |c | c |}
			\hline
			Cases&Vector ops. combination & Scalar and tensor ops. combination \\ 
			\hline
			\multirow{2}{*}{I}&$\frac{C^{(1)32}_{lq}}{\Lambda^2}=-9.96 \times 10^{-10}~ \rm GeV^{-2} $, & $\frac{C^{(1)32}_{lequ}}{\Lambda^2}=-9.27 \times 10^{-10}~ \rm GeV^{-2} $,  \\  
			&$\frac{C^{32}_{qe}}{\Lambda^2}=0$  & $\frac{C^{(3)32}_{lequ}}{\Lambda^2}=0$  \\  
			\hline
			\multirow{2}{*}{II}&$\frac{C^{(1)32}_{lq}}{\Lambda^2}=0$,  &$\frac{C^{(1)32}_{lequ}}{\Lambda^2}=0 $  \\ 
			&$\frac{C^{32}_{qe}}{\Lambda^2}=-1.04 \times 10^{-9}~\rm GeV^{-2}$&$\frac{C^{(3)32}_{lequ}}{\Lambda^2}=7.24 \times 10^{-11}~ \rm GeV^{-2}$\\
			\hline
			\multirow{2}{*}{III}&$\frac{C^{(1)32}_{lq}}{\Lambda^2}=-9.96 \times 10^{-10}~\rm GeV^{-2}$,&$\frac{C^{(1)}_{lequ}}{\Lambda^2}=-9.27 \times 10^{-10}~ \rm GeV^{-2} $,\\
			&$\frac{C^{32}_{qe}}{\Lambda^2}=-1.04 \times 10^{-9}~\rm GeV^{-2}$ & $\frac{C^{(3)32}_{lequ}}{\Lambda^2}=7.24 \times 10^{-11}~ \rm GeV^{-2} $ \\ 
			\hline 
	\end{tabular}}
	\caption{Different benchmark cases of coupling combinations to study optimal correlations and separability from the SM.}
	\label{tab:cases}  
\end{table}

\begin{figure}[htb!]
	$$
	\includegraphics[height=6cm,width=7.5cm]{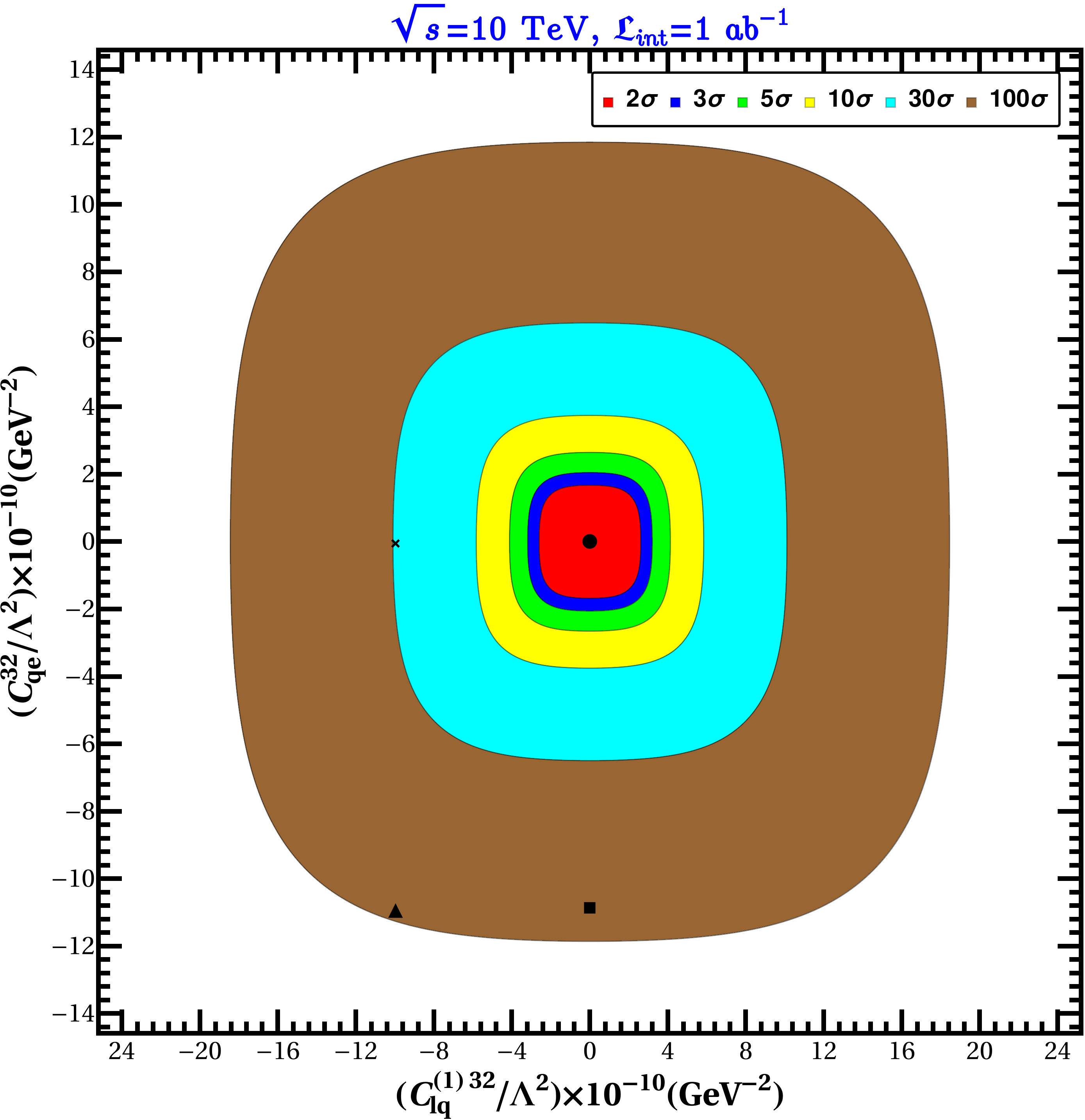}~~
	\includegraphics[height=6cm,width=7.5cm]{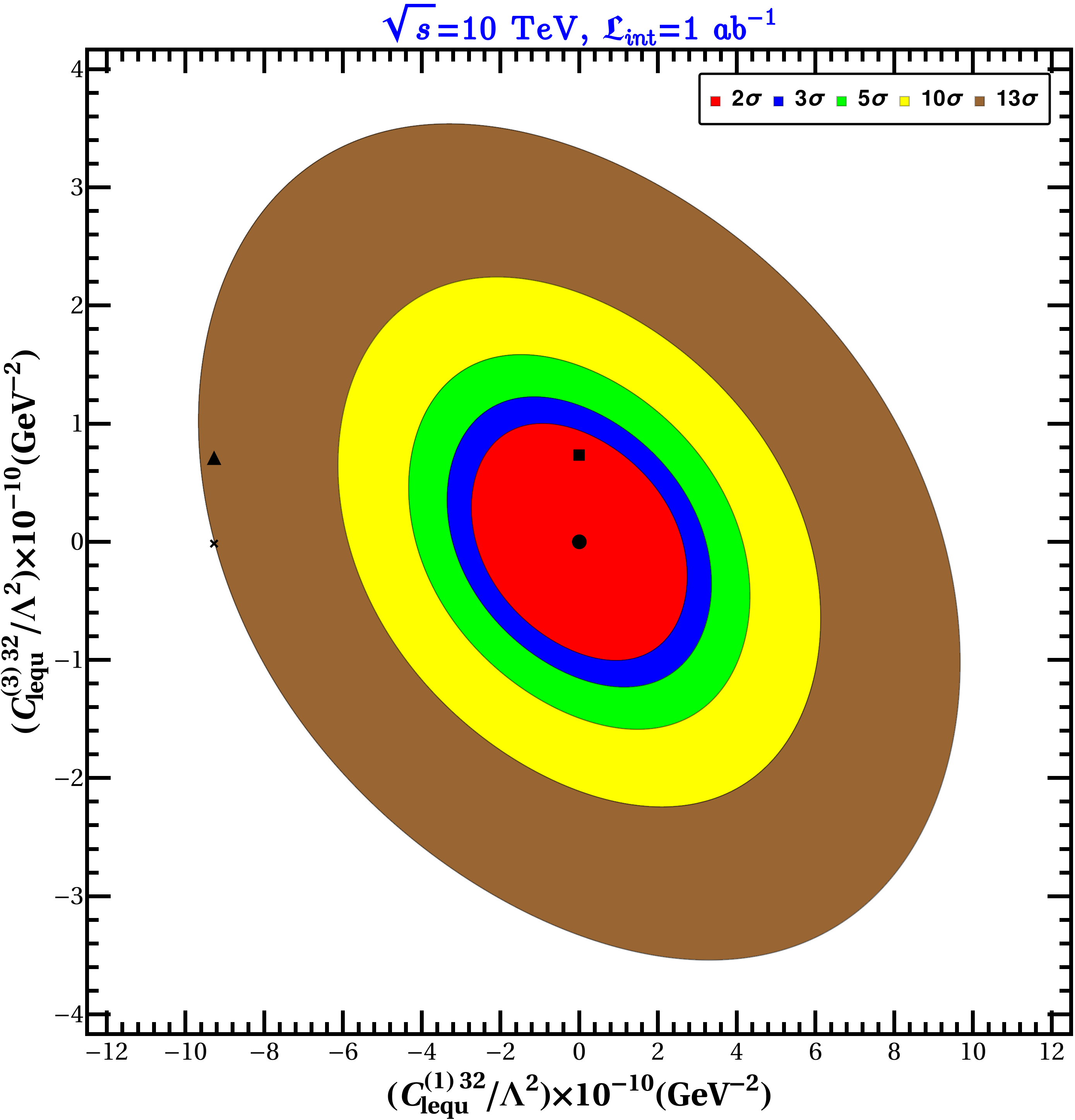}
	$$
	\caption{Separation of different cases mentioned in Table~\ref{tab:cases}; left: $C^{(1)32}_{lq}/\Lambda^2-C^{32}_{qe}/\Lambda^2$ plane, right: $C^{(1)32}_{lequ}/\Lambda^2-C^{(3)32}_{lequ}/\Lambda^2$ plane. The representative points related to case-I, case-II, and case-III are denoted by cross ($\times$), triangle ($\bigtriangleup$), and square ($\square$). The `base model' SM is represented by the dot ($\bullet$). Different $n \sigma$ regions are presented in the inset of the figures.}
	\label{fig:sig}
\end{figure}

In figure~\ref{fig:sig}, we show the optimal statistical separation of different cases listed in Table~\ref{tab:cases} from the `base model' SM. Here, 
Case-I and Case-II are followed from flavor constraints noted in Table~\ref{tab:cplng.const} and \ref{tab:cplng.const2} and case-III is considered to 
show the correlations among different operators. We consider the CM energy  $\sqrt{s}$ = 10 TeV and integrated luminosity $\mathfrak{L}_{\tt int}$= 1 $\rm{ab^{-1}}$.
We determine that for various vector (scalar and tensor) operator combinations, Case-I, Case-II, and Case-III are 28.18$\sigma$
(24.93$\sigma$), 76.97$\sigma$ (1.18$\sigma$), and 87.34$\sigma$ (23$\sigma$) away from SM respectively, so that the distinction of vector 
operators is comparatively easier than the scalar and tensor operators once we adhere to the benchmark points respecting flavor constraints. Estimated values of vector, scalar, and tensor couplings are noted in Table~\ref{tab:5sigma} for different CM energies and luminosities at 5$\sigma$ separation. To achieve the 5$\sigma$ separation for tensor operator with current flavor bound, we require 18 $\rm{ab^{-1}}$ integrated luminosity for 10 TeV CM energy.

\begin{table}[h]
	\centering
		\begin{tabular}{ |c |c | c | c | c| c|}
			\hline
			Couplings&$\sqrt{s}=10$ TeV, & $\sqrt{s}=10$ TeV,& $\sqrt{s}=30$ TeV, & $\sqrt{s}=30$ TeV, \\ 
			($\rm{GeV^{-2}}$)&$\mathfrak{L}_{\tt int}$=1 $\rm{ab^{-1}}$&$\mathfrak{L}_{\tt int}$=10 $\rm{ab^{-1}}$&$\mathfrak{L}_{\tt int}$=1 $\rm{ab^{-1}}$&$\mathfrak{L}_{\tt int}$=10 $\rm{ab^{-1}}$\\
			\hline
			$C^{(1)32}_{lq}/\Lambda^2$&$-4.13 \times 10^{-10}$&$-3.14 \times 10^{-10}$&$-7.92 \times 10^{-11}$&$-4.45 \times 10^{-11}$\\
			\hline 
			$C^{(1)32}_{lequ}/\Lambda^2$&$-7.88 \times 10^{-11}$&$-2.33 \times 10^{-11}$&$-7.88 \times 10^{-11}$&$-4.43 \times 10^{-11}$\\
			\hline
		    $C^{(3)32}_{lequ}/\Lambda^2$&$14.93 \times 10^{-11}$&$8.39 \times 10^{-11}$&$2.81 \times 10^{-11}$&$1.58 \times 10^{-11}$\\
		    \hline
	\end{tabular}
	\caption{Values of  vector ($ C_{lq}^{(1)32}/\Lambda^{2}$), scalar ($ C_{lequ}^{(1)32}/\Lambda^{2}$), and tensor ($ C_{lequ}^{(3)32}/\Lambda^{2}$) couplings at different CM energies and luminosities at 5$\sigma$ separation.}
	\label{tab:5sigma} 
\end{table}

\section{Sensitivity of "flavor-relaxed" $\left(C_{lequ}^{(1)32}/\Lambda^{2}\right)$ at the 500 GeV ILC}
\label{sec:scalar}
As seen from Table \ref{tab:cplng.const2}, the simultaneous fit on the scalar-mediated WC from $3 \to 2$ transition observables provides a less stringent 
bound compared to the other cases or when we combine all the flavor-violating observables together. Specifically, the order of $C_{lequ}^{(1)32}/\Lambda^{2}$ from 
$3 \to 2$ transition observables is $\sim 10^{-7}$ GeV$^{-2}$, whereas the best fit value from combining all the flavor-violating observables is of the order 
$\sim 10^{-9}$ GeV $^{-2}$. Therefore, $C_{lequ}^{(1)32}/\Lambda^{2}$ can be probed at a much lower CM energy preferable at the electron-positron machine 
such as International Linear Collider (ILC) \cite{Behnke:2013xla,ILCInternationalDevelopmentTeam:2022izu}. In this segment, we provide analyse probing this 
operator at the ILC with $\sqrt{s}=500$ GeV. At this CM energy, the outgoing particles are less boosted and hence collimation of decay products, as observed at 
multi-TeV colliders, is less likely to occur. Hence, an untagged di-jet analysis in such scenario will not be much helpful to segregate the signal processes from the huge 
two jet background possibilities. This motivates us to revert back to the traditional channels emerging from the heavy particle decay.

\begin{figure}[!htbp]
	\centering
	\includegraphics[width= 4.9cm, height=4.5cm]{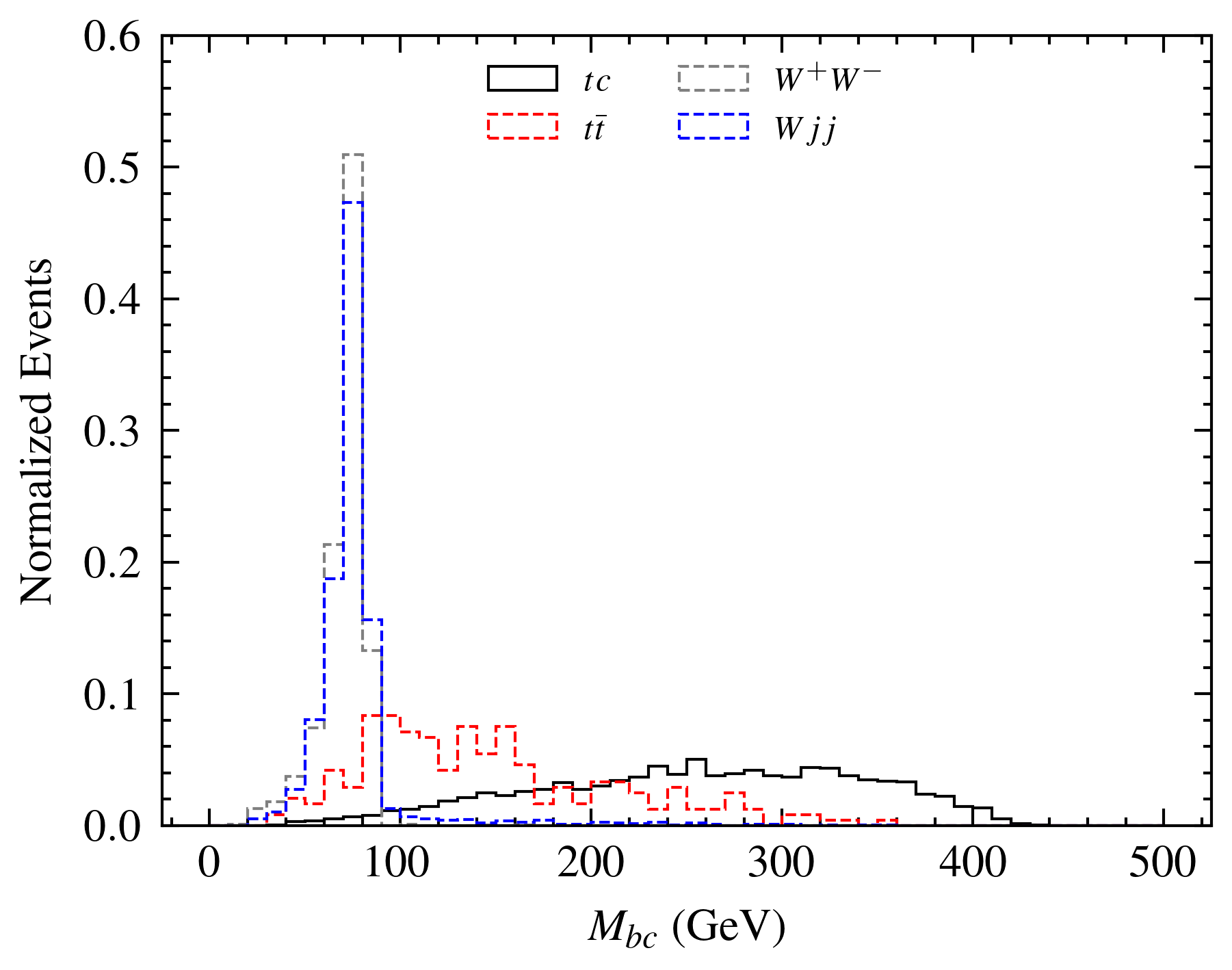}
	\includegraphics[width= 4.9cm, height=4.5cm]{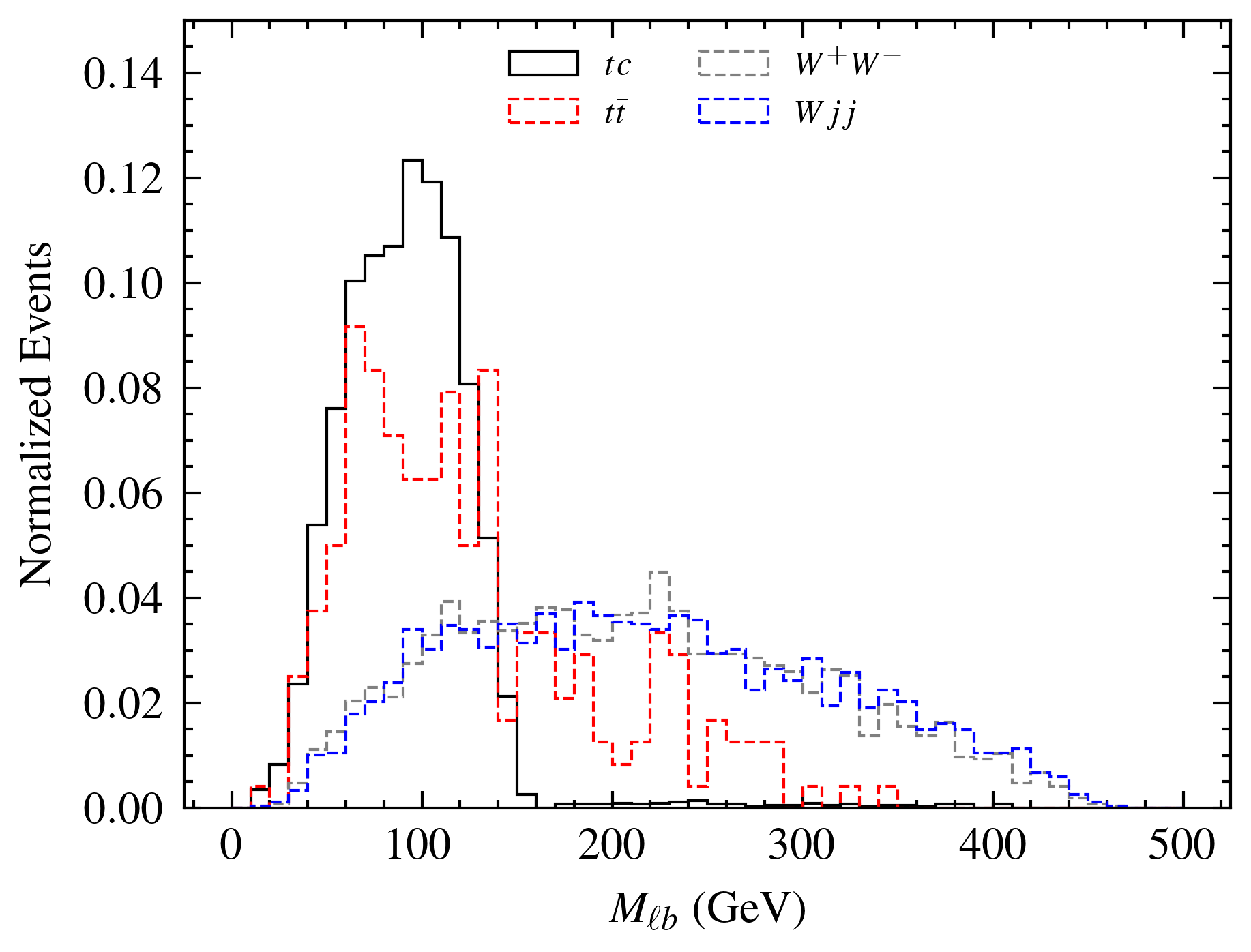}
	\includegraphics[width= 4.9cm, height=4.5cm]{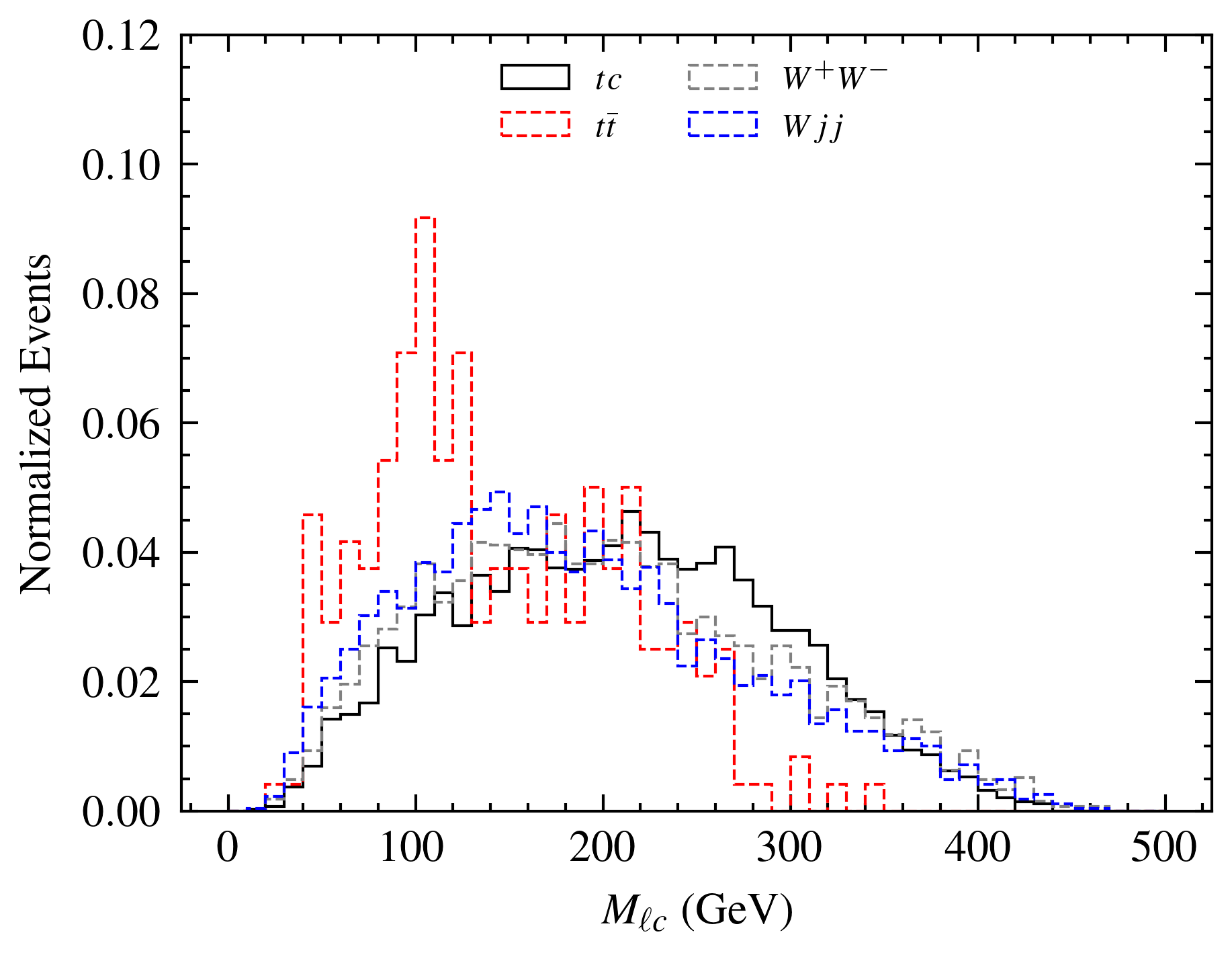}
	\caption{Invariant mass distributions corresponding to $\ell b c$ signal for signal and background processes at ILC 500 GeV run. For the signal process, we choose the benchmark: $C_{lequ}^{(1)32}/\Lambda^{2} = 1.79 \times 10^{-7}~\rm{GeV^{-2}}$, based on the flavor constraint obtained from Table~\ref{tab:cplng.const2}.}
	\label{fig:ilc1}
\end{figure}

For $tc$ production, we choose the signal process $t(bW(\ell \nu)) c$ $i.e.$ di-jet (one $b$-tagged, other $c$-tagged) plus one lepton with missing energy 
($\slashed{E}$). The relevant non-interfering SM background processes are $W(\ell \nu)W(jj)$, $t(bW(\ell \nu))\overline{t}(\overline{b}W(\ell \nu))$ and $W(\ell \nu)jj$. 
Here, $j$ corresponds to light jets ($u, d, s$), $c$-jets and $b$-jets collectively. Due to the large uncertainty in $c$ and $b$ tagging, miss-tagging of light jets and 
$c$-jets as $b$-jets and light jets and $b$-jets as $c$-jets should be taken into account. The $t\overline{t}$ background contributes as a result of a missing lepton 
and miss-tagging of $b$ from top decay as $c$. The event simulation is done in similar lines with Section~\ref{sec:es}. The $b$ and $c$ tagging are done using 
\textit{ILCgen} \texttt{Delphes} card based on ILC Snowmass projection \cite{ILCInternationalDevelopmentTeam:2022izu}. The $b$ and $c$ tagging efficiencies are 
0.702 and 0.315 respectively\footnote{The mistagging rates of $b(c)$ jet as light jets and $c(b)$ jet are 0.0613(0.0168) and 0.289(0.105) respectively.}. The invariant 
mass of the outgoing products turn out to be an important discriminator of the signal and background process, as shown in figure~\ref{fig:ilc1}. The jets ($b$ and $c$) in 
$WW$ and $Wjj$ are expected to peak at the $W$ pole which isn't the case for $tc$ and $t\overline{t}$. Similarly, the top decay products, $b \ell$ are expected to peak around 
100 GeV, taking into account the energy carried away by the invisible neutrino. Finally the $c$-jet from the signal process results from a contact interaction, in contrary to 
$WW$ and $t\overline{t}$, where it is produced in $W$ decay; thus the invariant mass between $c$ and $\ell / b$ for the signal is expected to be shifted more towards the 
CM energy, compared to the backgrounds. This works as an important discriminant in segregating the signal from $t \overline{t}$ background. The event cross-section 
after subsequent cutflow is shown in Table~\ref{tab:scalar.tab}. After all the kinematical cuts, 5$\sigma$ significance for this specific final state signal can be achieved 
at 10 fb$^{-1}$ luminosity with $\sqrt{s}=500$ GeV at the ILC.

\begin{table}[!htbp]
	\centering
	\begin{tabular}{|c|c|c|c|c|}
		\hline
		\multirow{2}*{Processes} & $\sigma(\mathcal{S})$  &  $\sigma(\mathcal{C}_{1})$  & $\sigma(\mathcal{C}_{2})$ & $\sigma(\mathcal{C}_{3})$ \\
		& $(N_{\ell}, N_{b}, N_{c}) = (1, 1, 1)$ & $M_{bc} > 100$ GeV & $M_{\ell b} < 150$ GeV & $M_{\ell c} > 150$ GeV \\
		\hline
		$tc$ & 6296 & 6057 & 5975 & 4413 \\
		$t \overline{t}$ & 1366 & 979 & 723 & 330 \\
		$WW$ & 18899 & 28 & 7 & 0 \\
		$Wjj$ & 17317 & 821 & 291 & 110 \\
		\hline
	\end{tabular}
	\caption{Signal and background events following each cut ($\mathcal{S}$, $\mathcal{C}_{1}$, $\mathcal{C}_{2}$, $\mathcal{C}_{3}$) at the ILC with 500 GeV 
	CM energy and 1 ab$^{-1}$ luminosity. Here, $N_{\ell}$, $N_{b}$ and $N_{c}$ refer to the number of leptons, $b$-tagged jets and $c$-tagged jets respectively. 
	$M_{ij}$ is the invariant mass of the $i^{th}$ and $j^{th}$ particles. For the signal process, we choose benchmark value 
	$C_{lequ}^{(1)32}/\Lambda^{2} = 1.79 \times 10^{-7}$ GeV$^{-2}$.}
	\label{tab:scalar.tab}
\end{table}

\begin{figure}[!htbp]
	\centering
	$$
	\includegraphics[scale=0.2]{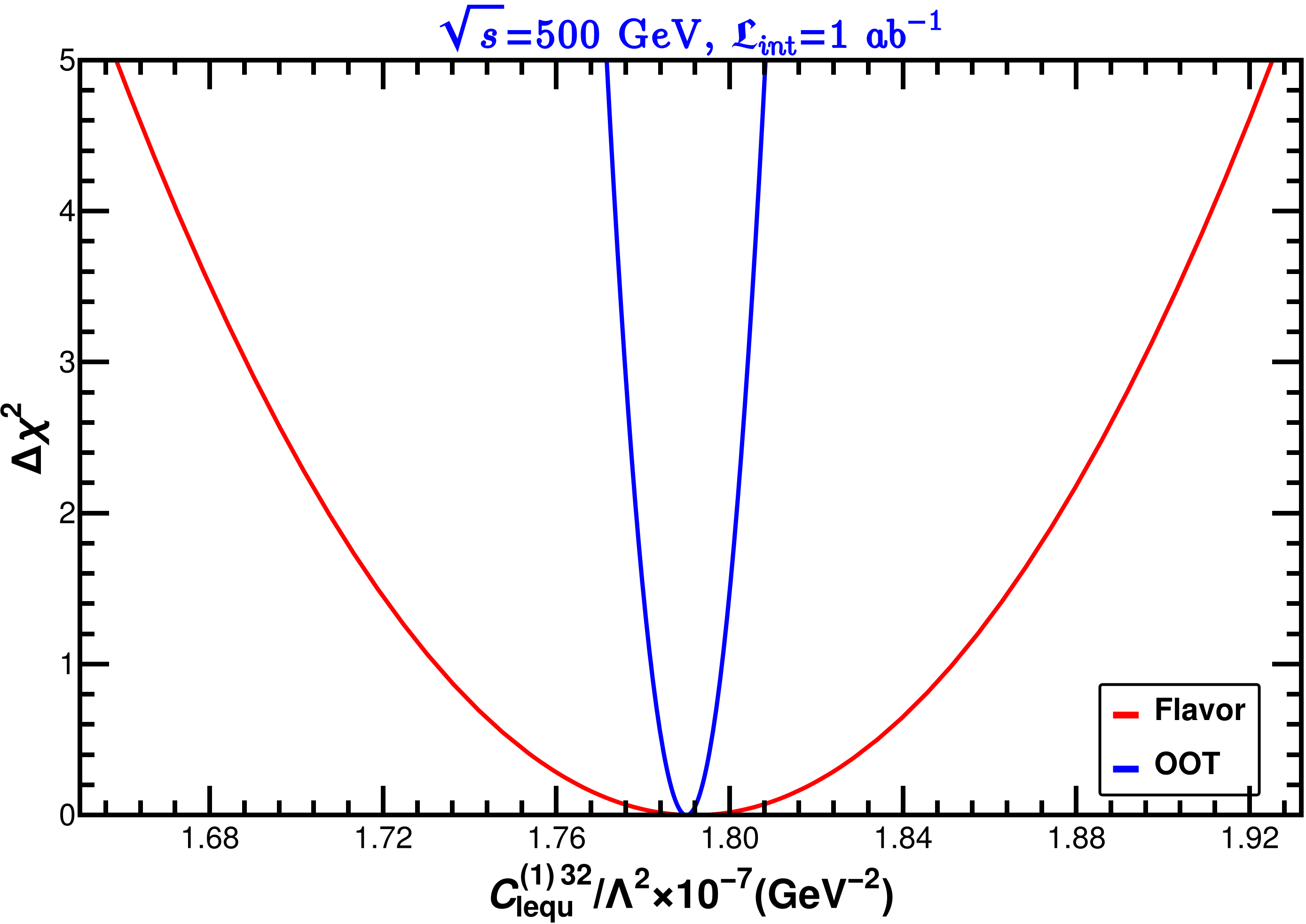}~~
	$$
	\caption{$\Delta \chi^2$ variation in red (blue) for flavor (OOT) as a function of dimension-six scalar mediated effective coupling at benchmark 
	$C_{lequ}^{(1)32}/\Lambda^{2} = 1.79 \times 10^{-7}~\rm{GeV^{-2}}$ based on the flavor constraint obtained from Table~\ref{tab:cplng.const2}.}
	\label{fig:chi.sqr.ilc}
\end{figure}

Subsequently, we analyze the optimal sensitivity of $C_{lequ}^{(1)32}/\Lambda^{2}$ with $\sqrt{s}=500$ GeV and $\mathfrak{L}_{\tt int}$=1 ab$^{-1}$ at the ILC. 
The optimal $\Delta \chi^2$ variation with $C_{lequ}^{(1)32}/\Lambda^{2}$ is shown in the figure~\ref{fig:chi.sqr.ilc} and compared with the flavor data. The figure 
demonstrates that the optimal uncertainty is much tighter than the flavor uncertainty in this case.

\section{Summary and Conclusion}
\label{sec:conclude}

In this paper, we have explored the SMEFT operators that contribute to $tc ~(\bar{t}c+t\bar{c})$ production at colliders. The most significant contribution arises 
from the quark flavor violating four-Fermi operators. The other operator which provides modification to $Ztc$ vertex has much less contribution at high 
CM energy and have thus been ignored in this analysis. Such four-Fermi operators contribute significantly to low energy flavor dependent processes and 
are thus heavily constrained by them. We study all such FCNC and FCCC processes including $b\to s(d)$ transitions, $P\to \ell\nu$, $K_L \to \mu\mu$, 
$B-\bar{B}$ mixing, top decays etc., to constrain the four-Fermi SMEFT operators. The most stringent bound arises on the tensor operator, having 
$C^{(3)32}_{lequ}/\Lambda^2 \sim 10^{-11}~ \rm GeV^{-2} $, followed by a conservative limit on scalar operator $C^{(1)32}_{lequ}/\Lambda^2 \sim 10^{-9}~ \rm GeV^{-2} $, and 
vector operator $C^{(1)32}_{lq}/\Lambda^2\sim 10^{-9}~ \rm GeV^{-2} $ (Although, the order is same, scalar operators is slightly more constrained than vector operator).  
We also predict observational sensitivities of the processes like $K_L \to \pi_0 \ell \ell$, $D_0 \to \mu\mu$, $t \to c\ell\ell$ and $t \to c \gamma$ from the obtained limits and 
show that they are consistent with the existing upper bounds. 


Using these constraints, we have examined the future probe of these SMEFT operators at multi-TeV moun collider through $tc ~(\bar{t}c+t\bar{c})$ 
production. The required CM energy is 10 TeV where the production cross-section is of the order of fb, respecting EFT constraint $\Lambda>\sqrt{s}$. 
Given the high CM energy, the process basically yields di-jet final state (a `top' jet and a `charm' jet) with no leptons stemming from the top decay. 
There is apparently a little chance of tagging them as well, which incorporates several SM backgrounds. There exists very little number of variables 
at disposal to segregate the signal from background, amongst which invariant di-jet mass and invariant mass of the heavy jet plays an important role 
to achieve a satisfactory signal significance for vector and scalar operator benchmarks in particular, at high luminosity 1 $\rm{ab^{-1}}$.


Using OOT, we have determined the optimal uncertainties of the vector, scalar and tensor type effective couplings at the benchmarks respecting flavor constraints. 
We see that at 10 TeV CM energy muon collider with 1 $\rm{ab^{-1}}$ integrated luminosity optimal uncertainty of the effective $C/\Lambda^2$ is better than 
the flavor uncertainty. Relative contribution between NP signal and non-interfering SM background play an important role in NP estimation, the less the 
background, the better the estimation. The dependence of the uncertainty on the CM energy and luminosity has also been compared and the advantages of 
CM energy to estimate optimal uncertainty pertaining to the EFT limit is discussed. Considering SM as a base model, distinction of different operators from the
SM has also been studied. We have observed that at 10 TeV CM energy, 30 $\rm{fb}^{-1}$ luminosity is required to segregate (at 5$\sigma$ level) vector 
($\mathcal{O}^{(1)}_{lq}$) operators from the SM, whereas for scalar ($\mathcal{O}^{(1)}_{lequ}$) and tensor ($\mathcal{O}^{(3)}_{lequ}$) operators we 
need  40 $\rm{fb^{-1}}$ and 18 $\rm{ab^{-1}}$ integrated luminosity, respectively, after obeying flavor constraints. However, in a more optimistic 
case of less stringent scalar operator having $C^{(1)32}_{lequ}/\Lambda^2 \sim 10^{-7}$ GeV$^{-2}$, 5$\sigma$ segregation can be achieved at 500 GeV CM energy 
and 10 fb$^{-1}$ luminosity at the ILC.

\acknowledgments
Sahabub Jahedi would like to thank Lipika Kolay and Ipsita Saha for useful discussions.

\appendix

\section{Loop function for $B_q-\bar{B}_q$ mixings and amplitudes of $t \to b \ell \nu$}
\vspace{0.1cm}

\begin{itemize}
	
	\item Loop function:
	\beq
	\rm{\tt DiscB(a,b,c)}=\frac{1}{a}\lambda(a,b^2,c^2)Log\left[\frac{-a+b^2+c^2+\sqrt{\lambda(a,b^2,c^2)}}
	{2bc}\right],
	\eeq
	where $\lambda(x,y,z)=x^2+y^2+z^2-2(xy+yz+zx)$.
	
	\item \underline{$ t  \to b \ell \nu_{\ell} $ decay}:  The SM and BSM contributions to the $t (p) \to b (k) \ell (k_1) \nu(k_2)$ decay amplitude are written as
	\begin{align}
	\begin{split}
	\mathcal{M}_{\tt SM}^{t}=& \frac{g^2}{2\sqrt 2((q^2-m_W^2)+i \Gamma_W m_W)} \bar{u}(k) \gamma^{\mu} P_L u(p)\bar{u}(k_2) \gamma_{\mu} P_L v(k_1),\\
	\mathcal{M}_{\tt NP}^{t}=& - \frac{C^{(3)32}_{l q}}{\Lambda^2} \bar{u}(k) \gamma^{\mu} P_L u(p)\bar{u}(k_2) \gamma_{\mu} P_L v(k_1)\\
	&+\frac{C^{(1)32}_{l e q u}}{\Lambda^2}\left(\bar{u}(k_1)  P_R v(k_2) \right)\left(\bar{v}(p_2)  P_R u(p_1)\right)\\
	&+\frac{C^{(3)32}_{l e q u}}{\Lambda^2}\left(\bar{u}(k_1) \sigma^{\mu \nu} P_R v(k_2) \right)\left(\bar{v}(p_2) \sigma_{\mu \nu} P_R u(p_1)\right).
	\end{split}
	\end{align}

\end{itemize}

\section{Renormalization Group (RG) Equations}
\noindent
	The renormalization group equations of the relevant SMEFT couplings are written as
	\cite{Jenkins:2013zja,Jenkins:2013wua,Alonso:2013hga,Celis:2017hod}
\begin{equation}
\frac{d\tilde{C}_{i}}{d \ln{\mu}} = \frac{1}{16 \pi^{2}} \beta_{i},
\end{equation}
where $\tilde{C}_{i} = \left \{\frac{C_{lq}^{(1)32}}{\Lambda^{2}}, \frac{C_{lequ}^{(1)32}}{\Lambda^{2}}, \frac{C_{lequ}^{(3)32}}{\Lambda^{2}} \right \} $ are the SMEFT coefficients and $\beta_{i} = \{\beta_{lq}^{(1)32}, \beta_{lequ}^{(1)32}, \beta_{lequ}^{(3)32} \}$ are the respective 1-loop $\beta$ functions of the SMEFT operators. The $\beta$ functions are:
\begin{equation}
\begin{split}
\beta_{lq}^{(1)32} &=  g'^{2} \frac{C_{lq}^{(1)32}}{\Lambda^{2}}+ \left[\gamma_{q}\right]_{33} \frac{C_{lq}^{(1)32}}{\Lambda^{2}} + \left[\gamma_{q}\right]_{33} \frac{C_{lq}^{(1)32}}{\Lambda^{2}} , \\
&= \left(g'^{2} + \frac{1}{2} y_{t}^{2} \right) \frac{C_{lq}^{(1)32}}{\Lambda^{2}}; \end{split}
\end{equation}
\begin{equation}
\begin{split}
\beta_{lequ}^{(1)32} &= - \left( \frac{11}{3} g'^{2} + 8 g_{S}^{2}\right) \frac{C_{lequ}^{(1)32}}{\Lambda^{2}} + \left[\gamma_{q}\right]_{33} \frac{C_{lequ}^{(1)32}}{\Lambda^{2}} +  \left[\gamma_{u}\right]_{33} \frac{C_{lequ}^{(1)}}{\Lambda^{2}}, \\
& = - \left( \frac{11}{3} g'^{2} + 8 g_{S}^{2} - \frac{1}{2} y_{t}^{2} \right) \frac{C_{lequ}^{(1)32}}{\Lambda^{2}}; \\
\end{split}
\end{equation}
\begin{equation}
\begin{split}
 \beta_{lequ}^{(3)32} &= - \left( \frac{2}{9} g'^{2} - 3 g^{2} + \frac{8}{3} g_{S}^{2}\right) \frac{C_{lequ}^{(3)32}}{\Lambda^{2}} + \left[\gamma_{q}\right]_{33} \frac{C_{lequ}^{(3)32}}{\Lambda^{2}} +  \left[\gamma_{u}\right]_{33} \frac{C_{lequ}^{(3)32}}{\Lambda^{2}}, \\
& = - \left( \frac{2}{9} g'^{2} - 3 g^{2} + \frac{8}{3} g_{S}^{2} - \frac{1}{2} y_{t}^{2} \right) \frac{C_{lequ}^{(3)32}}{\Lambda^{2}}; \\
\end{split}
\end{equation}
Here, $\gamma_{q} = \frac{1}{2} \left(y_{u} y_{u}^{\dagger} + y_{d} y_{d}^{\dagger}\right)$ and $\gamma_{u} = \left(y_{u} y_{u}^{\dagger}\right)$. We consider, $y_{u} = y_{d} = 0$ except for $y_{t}$. The RGEs for the SM parameters are noted below:
\begin{equation}
\begin{split}
\frac{dg}{d \ln{\mu}} & = \frac{1}{16 \pi^{2}} \left( -\frac{19}{6} g^{3} \right), \\
\frac{dg'}{d \ln{\mu}} & = \frac{1}{16 \pi^{2}} \left( \frac{41}{6} g'^{3} \right), \\
\frac{dg_{S}}{d \ln{\mu}} & = \frac{1}{16 \pi^{2}} \left( -7 g_{S}^{2} \right), \\
\frac{d y_{t}}{d \ln{\mu}} & = \frac{y_{t}}{16 \pi^{2}} \left( \frac{9}{4} g^{2} - \frac{17}{12} g'^{2} - 8 g_{S}^{2} + \frac{9}{2} y_{t}^{2} \right). \\
\end{split}
\end{equation}

\begin{figure}[!htbp]
	\centering
	\includegraphics[width= 7.5 cm, height= 5 cm]{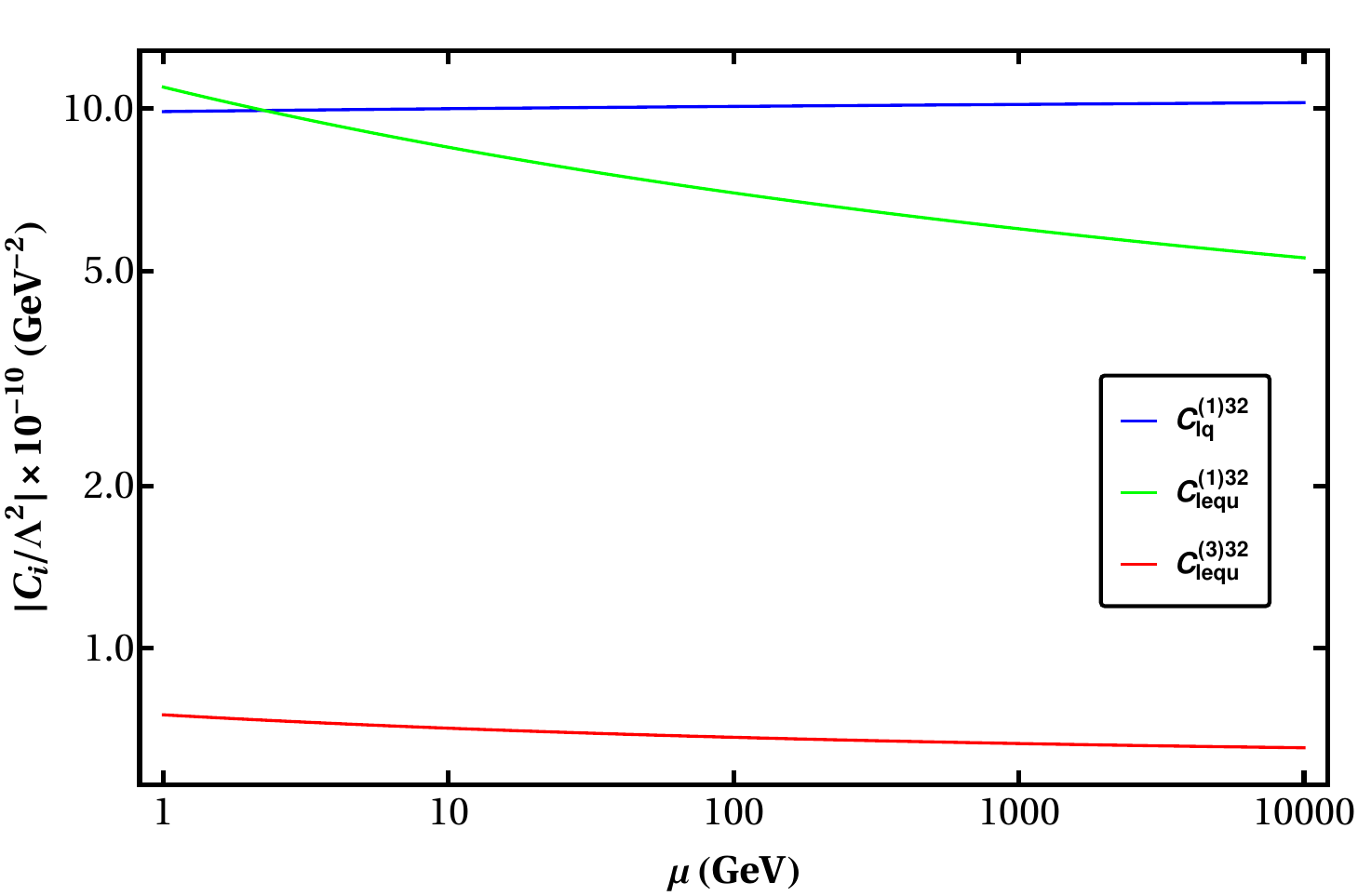}
	\caption{RG running for dimension-six vector ($ C_{lq}^{(1)32}/\Lambda^{2}$), scalar ($ C_{lequ}^{(1)32}/\Lambda^{2}$), and tensor ($ C_{lequ}^{(3)32}/\Lambda^{2}$) couplings with the energy scale ($\mu$).}
	\label{fig:rg}
\end{figure}

The variation of the dimension-six effective couplings with the renormalization scale ($\mu$) is illustrated in figure~\ref{fig:rg}, with comparisons and validation provided by {\tt Wilson}  \cite{Aebischer:2018bkb}.

\bibliographystyle{JHEP}
\bibliography{ref-flavour}
\end{document}